\pdfoutput=1
\PassOptionsToPackage{unicode}{hyperref}
\PassOptionsToPackage{hyphens}{url}
\PassOptionsToPackage{dvipsnames,svgnames,x11names}{xcolor}
\documentclass[
]{article}
\usepackage{xcolor}
\usepackage{amsmath,amssymb}
\usepackage{iftex}
\ifPDFTeX
  \usepackage[T1]{fontenc}
  \usepackage[utf8]{inputenc}
  \usepackage{textcomp} 
\else 
  \usepackage{unicode-math} 
  \defaultfontfeatures{Scale=MatchLowercase}
  \defaultfontfeatures[\rmfamily]{Ligatures=TeX,Scale=1}
\fi
\usepackage{lmodern}
\ifPDFTeX\else
\fi
\IfFileExists{upquote.sty}{\usepackage{upquote}}{}
\IfFileExists{microtype.sty}{
  \usepackage[]{microtype}
  \UseMicrotypeSet[protrusion]{basicmath} 
}{}
\makeatletter
\@ifundefined{KOMAClassName}{
  \IfFileExists{parskip.sty}{%
    \usepackage{parskip}
  }{
    \setlength{\parindent}{0pt}
    \setlength{\parskip}{6pt plus 2pt minus 1pt}}
}{
  \KOMAoptions{parskip=half}}
\makeatother
\makeatletter
\ifx\paragraph\undefined\else
  \let\oldparagraph\paragraph
  \renewcommand{\paragraph}{
    \@ifstar
      \xxxParagraphStar
      \xxxParagraphNoStar
  }
  \newcommand{\xxxParagraphStar}[1]{\oldparagraph*{#1}\mbox{}}
  \newcommand{\xxxParagraphNoStar}[1]{\oldparagraph{#1}\mbox{}}
\fi
\ifx\subparagraph\undefined\else
  \let\oldsubparagraph\subparagraph
  \renewcommand{\subparagraph}{
    \@ifstar
      \xxxSubParagraphStar
      \xxxSubParagraphNoStar
  }
  \newcommand{\xxxSubParagraphStar}[1]{\oldsubparagraph*{#1}\mbox{}}
  \newcommand{\xxxSubParagraphNoStar}[1]{\oldsubparagraph{#1}\mbox{}}
\fi
\makeatother

\usepackage{longtable,booktabs,array}
\usepackage{calc} 
\usepackage{etoolbox}
\makeatletter
\patchcmd\longtable{\par}{\if@noskipsec\mbox{}\fi\par}{}{}
\makeatother
\IfFileExists{footnotehyper.sty}{\usepackage{footnotehyper}}{\usepackage{footnote}}
\makesavenoteenv{longtable}
\usepackage{graphicx}
\makeatletter
\newsavebox\pandoc@box
\newcommand*\pandocbounded[1]{
  \sbox\pandoc@box{#1}%
  \Gscale@div\@tempa{\textheight}{\dimexpr\ht\pandoc@box+\dp\pandoc@box\relax}%
  \Gscale@div\@tempb{\linewidth}{\wd\pandoc@box}%
  \ifdim\@tempb\p@<\@tempa\p@\let\@tempa\@tempb\fi
  \ifdim\@tempa\p@<\p@\scalebox{\@tempa}{\usebox\pandoc@box}%
  \else\usebox{\pandoc@box}%
  \fi%
}
\def\fps@figure{htbp}
\makeatother

\NewDocumentCommand\citeproctext{}{}

\makeatletter
 \let\@cite@ofmt\@firstofone
 \def\@biblabel#1{}
 \def\@cite#1#2{{#1\if@tempswa , #2\fi}}
\makeatother
\newlength{\cslhangindent}
\newlength{\csllabelwidth}
\newenvironment{CSLReferences}[2] 
 {\begin{list}{}{%
  \setlength{\itemindent}{0pt}
  \setlength{\leftmargin}{0pt}
  \setlength{\parsep}{0pt}
  \ifodd #1
   \setlength{\leftmargin}{\cslhangindent}
   \setlength{\itemindent}{-1\cslhangindent}
  \fi
  \setlength{\itemsep}{#2\baselineskip}}}
 {\end{list}}
\usepackage{calc}

\providecommand{\tightlist}{%
  \setlength{\itemsep}{0pt}\setlength{\parskip}{0pt}}

\pdfoutput=1
\usepackage{float}
\usepackage{lmodern}
\usepackage{arxiv}
\usepackage{orcidlink}
\usepackage{amsmath}
\usepackage[T1]{fontenc}
\makeatletter
\@ifpackageloaded{caption}{}{\usepackage{caption}}
\AtBeginDocument{%
\ifdefined\contentsname
  \renewcommand*\contentsname{Table of contents}
\else
  \newcommand\contentsname{Table of contents}
\fi
\ifdefined\listfigurename
  \renewcommand*\listfigurename{List of Figures}
\else
  \newcommand\listfigurename{List of Figures}
\fi
\ifdefined\listtablename
  \renewcommand*\listtablename{List of Tables}
\else
  \newcommand\listtablename{List of Tables}
\fi
\ifdefined\figurename
  \renewcommand*\figurename{Figure}
\else
  \newcommand\figurename{Figure}
\fi
\ifdefined\tablename
  \renewcommand*\tablename{Table}
\else
  \newcommand\tablename{Table}
\fi
}
\@ifpackageloaded{float}{}{\usepackage{float}}
\floatstyle{ruled}
\@ifundefined{c@chapter}{\newfloat{codelisting}{h}{lop}}{\newfloat{codelisting}{h}{lop}[chapter]}
\floatname{codelisting}{Listing}

\makeatother
\makeatletter
\@ifpackageloaded{caption}{}{\usepackage{caption}}
\@ifpackageloaded{subcaption}{}{\usepackage{subcaption}}
\makeatother
\usepackage{bookmark}
\IfFileExists{xurl.sty}{\usepackage{xurl}}{} 
\hypersetup{
  pdftitle={E-values for Adaptive Clinical Trials: Anytime-Valid Monitoring in Practice},
  pdfauthor={Alexandra Sokolova; Vadim Sokolov},
  pdfkeywords={adaptive clinical trials, interim monitoring, optional
stopping, e-values, e-processes, group sequential
methods, alpha-spending, Bayesian adaptive designs, safe
testing, betting martingale, R package},
  colorlinks=true,
  linkcolor={blue},
  filecolor={Maroon},
  citecolor={Blue},
  urlcolor={Blue},
  pdfcreator={LaTeX via pandoc}}

\newcommand{\runninghead}{A Preprint }
\renewcommand{\runninghead}{E-values for Adaptive Clinical Trials }
\title{E-values for Adaptive Clinical Trials: Anytime-Valid Monitoring
in Practice}
\def\asep{\\\\\\ } 
\def\asep{\And }
\author{\textbf{Alexandra Sokolova}\\Knight Cancer Institute\\Oregon
Health and Science
University\\\\\href{mailto:sokolova@ohsu.edu}{sokolova@ohsu.edu}\asep\textbf{Vadim
Sokolov}\\Department of Systems Engineering and Operations
Research\\George Mason
University\\\\\href{mailto:vsokolov@gmu.edu}{vsokolov@gmu.edu}}
\date{}
\begin{document}
\maketitle
\begin{abstract}
Adaptive clinical trials increasingly rely on interim analyses, flexible
stopping, and data-dependent design modifications. These features can
improve ethics and efficiency but complicate statistical guarantees when
fixed-horizon test statistics are repeatedly inspected or reused after
adaptations. E-values and e-processes provide a practical alternative:
they yield anytime-valid tests and confidence sequences that remain
valid under optional stopping and optional continuation without
requiring a prespecified monitoring schedule.

This paper is a methodology guide for practitioners. We develop the
betting-martingale construction of e-processes for two-arm randomized
controlled trials, show how e-values naturally handle composite null
hypotheses and support futility monitoring, and provide guidance on when
e-values are appropriate, when established alternatives are preferable,
and how to integrate e-value monitoring with group sequential and
Bayesian adaptive workflows.

A numerical study compares five monitoring rules---naive and calibrated
versions of frequentist, Bayesian, and e-value approaches---in a two-arm
binary-endpoint trial. Naive repeated testing and naive posterior
thresholds inflate Type I error substantially under frequent interim
looks. Among the valid methods, the calibrated group sequential rule
achieves the highest power, the e-value rule provides robust
anytime-valid control with moderate power, and the calibrated Bayesian
rule is the most conservative.

Extended simulations show that the power gap between group sequential
and e-value methods depends on the monitoring schedule and reverses
under continuous monitoring. The methodology, including futility
monitoring, platform trial multiplicity control, and hybrid strategies
combining e-values with established methods, is implemented in the
open-source R package \texttt{evalinger} and situated within the
regulatory framework of the January 2026 FDA draft guidance on Bayesian
methodology.
\end{abstract}
{\bfseries \emph Keywords}
\def\sep{\textbullet\ }
adaptive clinical trials \sep interim monitoring \sep optional
stopping \sep e-values \sep e-processes \sep group sequential
methods \sep alpha-spending \sep Bayesian adaptive designs \sep safe
testing \sep betting martingale \sep 
R package

\newpage

\section{Introduction}\label{introduction}

Clinical trials are sequential in practice even when statistical
analysis plans are written as if inference occurs once, at a fixed
terminal sample size. Enrollment and outcome accrual occur over calendar
time; data monitoring committees review accumulating safety and efficacy
information; operational constraints and emerging external evidence can
precipitate protocol amendments. The practical question is not whether
interim decisions will be made, but how to make them without sacrificing
evidential reliability.

The need to quantify ``how stochastic'' a realized sequence looks has
appeared in domains far from clinical trials. Kolmogorov's work on
measuring the stochasticity of finite sequences, with applications in
areas including genetics and information theory, addressed settings
where observations arrive sequentially and the temptation to stop or
re-evaluate after seeing a striking pattern is ever-present (Ramdas et
al. 2023). Ville's 1939 martingale characterization of randomness, which
preceded Wald's sequential probability ratio test by several years,
established the key inequality: if a bettor's capital is a nonnegative
supermartingale, the probability that it ever exceeds \(1/\alpha\) is at
most \(\alpha\) (Ville 1939; Wald 1947). Shafer and Vovk later recast
this as the explicit ``testing by betting'' framework in a
game-theoretic probability setting, showing that it provides a
foundation for sequential inference that does not require a fixed sample
size or a prespecified stopping rule (Shafer and Vovk 2019; Shafer
2021). Modern e-values are precisely the payoffs of such bets, and
e-processes are the wealth trajectories of strategies that bet against
the null sequentially (Polson, Sokolov, and Zantedeschi 2026). The
betting metaphor in testing has a parallel in trial design: Berry's
decision-analytic framing of clinical trials as bandit problems also
involves sequential wagering---here, the ``bet'' at each patient is the
treatment assignment, and the payoff is the patient's outcome (D. A.
Berry 2025). Thompson's 1933 proposal to randomize to each arm with
probability equal to the current belief that the arm is best (Thompson
1933) was an early adaptive allocation strategy, predating both Hill's
fixed randomization and Wald's sequential testing, and it shares with
e-values the property that the allocation (or bet) at each step is a
predictable function of accumulated evidence. For clinical trials, this
confluence of betting-based testing and betting-based allocation offers
a clean framework: one bets against the null at each patient outcome,
and the accumulated evidence remains interpretable regardless of how
many times one has looked.

In contemporary trials, classical frequentist inference accommodates
interim monitoring through group sequential designs and alpha-spending
functions, at the cost of prespecifying a monitoring framework (a set of
interim looks, or a spending function and information-time definition)
and limiting formal Type I error guarantees to that framework (Pocock
1977; O'Brien and Fleming 1979; Lan and DeMets 1983; Jennison and
Turnbull 2000). Bayesian adaptive designs embrace continuous learning
and decision-making via posterior and predictive probabilities, with
operating characteristics assessed by simulation and controlled through
conservative decision thresholds (Spiegelhalter, Abrams, and Myles 2004;
S. M. Berry et al. 2010; D. A. Berry 2004, 2025; U.S. Food and Drug
Administration 2025). Meurer, Lewis, and Berry (Meurer, Lewis, and Berry
2012) argue that adaptive designs partially remedy the ``therapeutic
misconception''---the common misunderstanding that trial participation
guarantees individually optimized care---by steering more patients
toward effective treatments as evidence accumulates.

Berry's work provides a practitioner-oriented anchor for what ``Bayesian
monitoring'' means in operations (D. A. Berry 1989, 2025). In Berry's
decision-analytic framework, the clinical trial is fundamentally a
bandit problem: treating patients enrolled in the trial is as important
as treating patients who will present after the trial, and the design
should seek to balance earning (effective treatment of the current
patient) with learning (information that improves future treatment
decisions) (D. A. Berry 2004; D. A. Berry and Fristedt 1985). This
framing leads naturally to response-adaptive randomization via Thompson
sampling (Thompson 1933), whereby the next patient is assigned to the
arm with probability proportional to the current posterior probability
that the arm is best---a strategy that predates modern e-values by
decades but shares their betting-theoretic ancestry. The operational
decision quantity in Berry's framework is not the raw posterior
probability of superiority but the Bayesian predictive probability of
trial success: the probability, integrating over the current posterior,
that the trial will reach a positive conclusion if enrollment continues.
This quantity drives stopping rules in CALGB 49907, which stopped at 633
patients (versus a planned maximum of 1800) when the predictive
probability of a meaningful answer exceeded 80\% (Muss et al. 2009), in
the ThermoCool AF device trial, which stopped at its first interim when
the predictive probability reached 99.9\% (D. A. Berry 2025), and in the
bandit-inspired platform I-SPY 2, where monthly predictive probability
assessments drove arm graduation, futility, and response-adaptive
randomization across 23 investigational therapies and 10 molecular
signatures (Barker et al. 2009; Park et al. 2016; D. A. Berry 2025). The
important caveat, well understood by Bayesian trialists but sometimes
underappreciated in practice, is that posterior or predictive threshold
rules do not automatically control frequentist Type I error under
optional stopping; calibration, typically via simulation at the design
stage, is required whenever such guarantees are part of the design
specification (U.S. Food and Drug Administration 2025, 2026).

E-values---nonnegative statistics with expectation at most one under the
null---and their sequential counterparts, e-processes, provide a
complementary perspective on evidence and error control. Their defining
property yields tests that are valid under optional stopping and
optional continuation without requiring a fixed sample size or a
prespecified monitoring schedule (Grünwald, Heide, and Koolen 2023;
Ramdas et al. 2023; Shafer 2021). This flexibility simplifies
monitoring: the same threshold \(E_t \ge 1/\alpha\) controls Type I
error under arbitrary peeking. At the same time, e-values introduce new
design degrees of freedom: the construction of a powerful e-process
depends on the choice of betting strategy, alternative model, or
prior-like tuning parameter, and misspecification can reduce power or
increase expected sample size (Grünwald et al. 2021; Martin 2025).
Moreover, the regulatory ecosystem is deeply habituated to \(p\)-values,
confidence intervals, and familywise error control in confirmatory
settings, and the translation from e-values to reporting conventions
requires careful explanation and, in some settings, additional
methodology.

This paper has four goals. First, it provides a practical, clinically
oriented account of e-values and e-processes as tools for interim
monitoring and adaptive trial conduct, including explicit constructions
for two-arm trials with binary and survival endpoints. Second, it gives
practitioner guidance on when e-values should be used, when established
alternatives are preferable, and how futility, multiplicity, and
regulatory requirements interact with e-value methodology. Third, it
presents a numerical demonstration comparing five monitoring rules in a
two-arm randomized trial: naive repeated \(p\)-values, an
O'Brien--Fleming-like group sequential boundary, a betting-martingale
e-process, and both naive and simulation-calibrated Bayesian posterior
thresholds. Among the valid methods, the calibrated group sequential
rule achieves 86.1\% power, the e-process 72.3\%, and the calibrated
Bayesian rule 68.8\%, while naive methods inflate Type I error to
13--15\% under 20 interim looks. Fourth, it provides the
\texttt{evalinger} R package and an accompanying interactive web
application that implement the full methodology---design calibration,
real-time monitoring, confidence sequences, futility analysis, platform
trial multiplicity control, and head-to-head comparison with group
sequential and Bayesian approaches---enabling practitioners to apply
e-value monitoring within the regulatory framework of the 2026 FDA draft
Bayesian guidance (U.S. Food and Drug Administration 2026).

To clarify the scope of this paper: Sections 4--5 review established
theory (group sequential methods, e-values, e-processes, Ville's
inequality, the betting-martingale construction, and the GROW-optimal
betting fraction), drawing on Ramdas et al. (2023), Grünwald, Heide, and
Koolen (2023), and Shafer (2021). The original contributions are: (i)
the five-method numerical comparison including both naive and calibrated
Bayesian rules alongside group sequential and e-value monitoring
(Section 7); (ii) the extended simulations quantifying the dependence of
the power gap on monitoring schedule, design parameters, and trial
scale, including the reversal under continuous monitoring (Sections
7.5--7.8); (iii) the futility monitoring framework via both confidence
sequences and reciprocal e-processes, with Monte Carlo evaluation
(Sections 6.3 and 7.7); (iv) the three-way comparison (classical,
Bayesian, e-value) on the Novick (1965) real-data illustration (Section
8); (v) the practitioner guidance and regulatory analysis situating
e-values within the 2026 FDA draft Bayesian guidance (Sections 2--3 and
9); and (vi) the \texttt{evalinger} R package and interactive web
application.

\section{Practitioner guidance: when to use e-values in clinical
trials}\label{practitioner-guidance-when-to-use-e-values-in-clinical-trials}

E-values are most useful when unplanned or frequent peeking is
operationally likely, or when optional continuation is a realistic
possibility. This includes settings with irregular interim looks (for
example, DSMB meetings scheduled in calendar time rather than at fixed
information fractions), continuous safety monitoring in which rapid
signal detection is important, adaptive protocols involving sample size
re-estimation, enrichment, arm dropping or adding, or response-adaptive
randomization, and program-level evidence accumulation across substudies
or sequential experiments in which the decision to run a subsequent
study depends on earlier outcomes (Grünwald, Heide, and Koolen 2023).
The REMAP-CAP/COVID platform trial illustrates this: designed as a
perpetual adaptive platform for community-acquired pneumonia, it pivoted
within two days of the WHO pandemic declaration to begin enrolling
COVID-19 patients, eventually randomizing over 24,000 patients across 66
interventions in 15 countries (Angus et al. 2020; D. A. Berry 2025). In
such a setting, the analysis schedule was dictated by the evolving
pandemic rather than a pre-specified statistical plan, and anytime-valid
inference would have been a natural choice for the evidential framework
of the trial.

Group sequential methods remain the default for many confirmatory
settings because they are well understood by regulators, optimized for
common test statistics, and achieve higher power than e-values at the
same Type I error level when the monitoring schedule is fixed. When
interim looks are fixed and prespecified (or governed by an agreed-upon
spending function), endpoints are standard, and power efficiency at a
planned information time is a primary objective, alpha-spending and
group sequential boundaries provide a mature and widely accepted
solution. In such settings e-values can be used as a secondary
evidential stream, but they are not automatically superior to an already
well-calibrated group sequential design. Our numerical study in Section
Section~\ref{sec-numerical} quantifies this: in a two-arm binary trial
with 20 equally spaced looks, the calibrated group sequential boundary
achieves 86.1\% power while the betting e-process achieves 72.3\%, a gap
of approximately 14 percentage points that reflects the price of anytime
validity.

Bayesian adaptive designs excel when the decision problem is explicitly
about probability of success, utility, or learning across strata or arms
(D. A. Berry 2025). In such cases, posterior and predictive
probabilities are a natural operational currency, and Berry's
decision-analytic framework provides a mature toolbox for design,
monitoring, and adaptation. However, as our numerical study confirms,
posterior-threshold stopping at a fixed level (e.g.,
\(\Pr(p_T > p_C \mid \text{data}) > 0.975\)) can inflate Type I error
dramatically under frequent interim looks. In our example with 20 looks,
the naive posterior-threshold rule has Type I error about 13.5\%. The
calibrated Bayesian rule requires a posterior threshold of approximately
0.998 to achieve 2.5\% Type I error with 20 looks---substantially more
conservative than the naive 0.975. This calibrated threshold achieves
68.8\% power (more conservative than the betting e-process in this
design), and it offers somewhat earlier stopping under the alternative
(average 134 patients per arm versus about 140 for both the calibrated
group sequential and e-value rules).

\subsection{Futility monitoring and integration with Bayesian
workflows}\label{futility-monitoring-and-integration-with-bayesian-workflows}

Futility---the decision to stop a trial because the treatment is
unlikely to demonstrate a meaningful benefit even if enrollment
continues---is as important as efficacy monitoring in adaptive trials.
E-processes support futility monitoring through two complementary
mechanisms. First, a confidence sequence for the treatment effect can be
computed at each interim look (see Section Section~\ref{sec-reporting}),
and futility can be declared when the upper bound of the confidence
sequence falls below the minimum clinically important difference.
Interpreted at the chosen error level, this provides an anytime-valid
signal that a clinically meaningful effect is not supported by the data
accumulated so far. Second, one can construct a ``reciprocal'' e-process
that tests the reverse hypothesis (the treatment is at least as good as
control by a clinically meaningful margin); if this reciprocal e-process
grows large, there is strong evidence against the possibility of a
meaningful treatment effect. In practice, the confidence-sequence
approach is more transparent for DSMB communication because it directly
conveys the range of plausible effect sizes, whereas the reciprocal
e-process provides a more formal test of futility that can be integrated
into a multiplicity-adjusted decision framework. Both mechanisms are
implemented in \texttt{evalinger} (\texttt{futility\_cs()} and
\texttt{futility\_eprocess()}), and the Monitoring Dashboard in the web
application provides live confidence-sequence visualization to support
real-time futility assessment.

D. A. Berry (2025) explicitly frames clinical trial design as
decision-making under uncertainty, where adapting based on accumulating
data is a feature rather than a violation, and where predictive
probabilities are a natural quantity for go/no-go and sample size
decisions. Berry's vision is that the ``observer'' driving adaptations
should be a computer armed with a prospective algorithm that dictates
adaptations determined in advance of the trial. This
principle---algorithmic, prospective adaptation---aligns directly with
the e-process requirement that betting fractions be predictable
functions of past data. In this perspective, e-values complement
Bayesian decision rules. When a trial uses predictive-probability
thresholds for operational actions---as in AWARD-5, which seamlessly
combined phases 2 and 3 using biweekly Bayesian analyses to select two
doses of dulaglutide from among seven candidates without pausing
enrollment (Geiger et al. 2012; D. A. Berry 2025)---an e-process can be
run in parallel as an auditable, anytime-valid evidential ledger for
Type I error control. This does not replace the Bayesian decision rule;
rather, it provides a conservative and transparent frequentist safety
layer for interim efficacy claims. The combination is natural for
seamless-phase designs like GBM AGILE, where an investigational arm's
Stage 1 (response-adaptive) and Stage 2 (confirmatory extension)
patients are pooled for a single registration analysis (Alexander et al.
2018; D. A. Berry 2025): an e-process accumulating evidence across both
stages would provide anytime-valid error control without requiring the
stages to have separate statistical analysis plans. When adaptation
logic is complex and heavily simulation-calibrated, e-values can also
simplify monitoring governance, because they offer a single time-uniform
threshold that remains valid as the schedule of looks changes, reducing
pressure for ad hoc boundary modifications.

\subsection{Software, regulatory landscape, and common
pitfalls}\label{software-regulatory-landscape-and-common-pitfalls}

Software for e-value construction in clinical trials is maturing. The
\texttt{safestats} R package (Schure and Ly 2022) implements safe tests
for common endpoints including the safe logrank test for survival data,
safe \(t\)-tests, and safe tests for proportions, with calibration tools
for power and expected sample size under design alternatives. The
broader e-value literature is supported by research software in R and
Python, though production-grade clinical trial software remains less
developed than for group sequential methods (e.g., \texttt{gsDesign},
\texttt{rpact}). From a regulatory standpoint, the U.S. Food and Drug
Administration's guidance on adaptive designs emphasizes that any
adaptive method must prospectively control Type I error and maintain
trial integrity (U.S. Food and Drug Administration 2019). E-values
satisfy this requirement by construction (Ville's inequality provides an
analytic guarantee), but regulatory familiarity with e-values is
limited. Communication strategies should emphasize the mapping from
e-values to familiar quantities: always-valid \(p\)-values
(\(p_t = 1/\sup_{s \le t} E_s\)), confidence sequences, and calibrated
effect estimates. In submissions, e-value monitoring can be presented as
a pre-planned sequential analysis method with a single time-uniform
threshold, analogous to group sequential monitoring but without the
constraint of a fixed look schedule. The FDA Bayesian guidance for
medical devices (U.S. Food and Drug Administration 2025) provides a
useful precedent for non-traditional evidential frameworks in regulatory
review, and the January 2026 CDER/CBER draft guidance on Bayesian
methodology for drugs and biologics (U.S. Food and Drug Administration
2026) substantially expands the regulatory framework by codifying three
pathways for success criteria---Type I error calibration, direct
posterior interpretation, and decision-theoretic approaches---any of
which can accommodate e-value monitoring.

Two recurrent pitfalls deserve emphasis. The first is repeatedly
monitoring a fixed-horizon \(p\)-value at level \(\alpha\) without
sequential correction, and the second is assuming that a posterior
probability threshold (for example, exceeding \(0.975\)) is
automatically safe under optional stopping; both can yield substantial
Type I error inflation, as our numerical study demonstrates (Section
Section~\ref{sec-numerical}). A third pitfall, specific to e-values, is
choosing a betting strategy or mixture alternative carelessly: an
e-process tuned for a treatment effect much larger or smaller than the
truth may be severely underpowered. Good practice requires documenting
the filtration explicitly, including what information was available at
each interim look and which adaptation decisions were permitted as
functions of the past. E-process construction should be treated as a
design choice requiring power evaluation, typically via simulation, in
the same spirit as boundary selection in group sequential design.

\section{Practical recipe: implementing an e-value monitoring
plan}\label{practical-recipe-implementing-an-e-value-monitoring-plan}

The implementation begins by specifying the decision statement,
including the null hypothesis \(H_0\), the estimand-aligned endpoint,
and the analysis population. The monitoring error budget is then
selected by fixing \(\alpha\) and deciding whether monitoring is
one-sided or two-sided; interim decisions are mapped to a threshold of
the form \(E_t \ge 1/\alpha\), with appropriate familywise or
multiplicity adjustments in multi-arm settings. Construction of an
e-process is endpoint-dependent. For two-arm trials with binary
outcomes, the betting martingale described in Section
Section~\ref{sec-betting} provides a direct construction (implemented by
\texttt{eprocess\_binary()} in the \texttt{evalinger} package). For
survival endpoints, the safe logrank test provides a principled option
(Grünwald et al. 2021) (implemented by \texttt{eprocess\_logrank()}).
When some prior information is available but robustness is desired,
regularized e-processes provide a tunable compromise between efficiency
and stability (Martin 2025).

Reporting should be prespecified, including whether the DSMB will review
\(E_t\), \(\log E_t\), or derived always-valid confidence sequences, and
what constitutes an actionable signal for efficacy, futility, or safety.
Operating characteristics should be evaluated under the null and under
clinically meaningful alternatives to quantify power, expected sample
size, and sensitivity to misspecification.

Unlike group sequential designs, where power calculations are based on
the information fraction and spending function, e-value-based designs
require reasoning about the expected log-growth rate of the e-process
under the alternative. For the betting martingale
\(E_n = \prod_{i=1}^n (1 + \lambda D_i)\), the expected stopping time
under a design alternative \((p_T, p_C)\) can be approximated by
\(\tau \approx \log(1/\alpha) / g(\lambda;\, p_T, p_C)\), where
\(g(\lambda;\, p_T, p_C) = \mathbb{E}_{H_1}[\log(1 + \lambda D_i)]\) is
the per-observation expected log-growth rate. Note that this rate
depends on the individual response rates \(p_T\) and \(p_C\), not just
their difference \(\delta = p_T - p_C\): two alternatives with the same
\(\delta\) but different base rates yield different growth rates,
because the probability of discordant outcomes (the only patient pairs
that contribute to evidence) depends on the base rates. For a fixed
design alternative, the growth rate is maximized by the
growth-rate-optimal (GROW) choice of \(\lambda\). In practice, designers
should compute \(g(\lambda;\, p_T, p_C)\) for a grid of \(\lambda\)
values and clinically plausible alternatives, select \(\lambda\) to
maximize power (or minimize expected sample size) under the design
alternative, and verify by simulation; the \texttt{evalinger} functions
\texttt{grow\_lambda()}, \texttt{grow\_lambda\_grid()}, and
\texttt{edesign\_binary()} automate this workflow. The maximum sample
size \(N_{\max}\) should be chosen so that the e-process has high
probability of crossing \(1/\alpha\) before \(N_{\max}\) under the
design alternative, with a recommended target of 80--90\% power. The
interactive web application bundled with \texttt{evalinger} provides a
Design Calculator that visualizes \(g(\lambda;\, p_T, p_C)\) and
expected stopping time as functions of the betting fraction, allowing
rapid exploration of the design space without custom simulation code.

\section{Background: group sequential monitoring in confirmatory
trials}\label{background-group-sequential-monitoring-in-confirmatory-trials}

Repeated inspection of a fixed-horizon test statistic inflates the
probability of a false positive unless the statistic and decision rule
are explicitly designed to be valid over time. This phenomenon motivates
sequential analysis beginning with Wald's SPRT, Robbins's sequential
design framework (Robbins 1952), and their clinical trial extensions
(Wald 1947; Armitage 1954, 1960). Group sequential methods
operationalize these ideas by partitioning information accrual into a
finite number of interim analyses and constructing boundaries so that
the overall Type I error rate is controlled (Pocock 1977; O'Brien and
Fleming 1979; Lan and DeMets 1983; Jennison and Turnbull 2000).

Group sequential tests typically assume that a sequence of test
statistics at interim looks follows a multivariate normal distribution
with a known correlation structure driven by information fractions.
Canonical boundaries include the Pocock and O'Brien--Fleming families.
Alpha-spending functions (Lan and DeMets 1983) generalize these
boundaries by allocating cumulative Type I error as a function of
information time, permitting flexible look schedules.

\section{E-values and e-processes: a
mini-tutorial}\label{e-values-and-e-processes-a-mini-tutorial}

This section provides a self-contained introduction to e-values and
e-processes for readers coming from a clinical trials background. For
comprehensive treatments, see Ramdas et al. (2023), Vovk and Wang
(2021), and Grünwald, Heide, and Koolen (2023).

\subsection{What is an e-value?}\label{what-is-an-e-value}

An \textbf{e-value} is a nonnegative statistic \(E \ge 0\) whose
expected value under the null hypothesis is at most 1: \[
\mathbb{E}_{H_0}[E] \le 1.
\] That is the entire definition. The ``e'' stands for ``evidence'' (the
intended use) or ``expectation'' (the defining constraint) (Vovk and
Wang 2021). To use an e-value as a test, reject \(H_0\) when
\(E \ge 1/\alpha\). This controls Type I error at level \(\alpha\) by
Markov's inequality: \[
\mathbb{P}_{H_0}\!\left(E \ge \frac{1}{\alpha}\right) \le \alpha.
\]

For a one-sided test at \(\alpha = 0.025\) (the standard in confirmatory
clinical trials), we reject when \(E \ge 40\). An e-value of 50 is
stronger evidence than an e-value of 25, just as a smaller \(p\)-value
is stronger evidence. In contrast to \(p\)-values, e-values admit simple
and \emph{valid} combination operations in many settings: under
independence (or more generally conditional validity given past
information), the product of e-values remains an e-value, and under
arbitrary dependence the arithmetic mean remains an e-value (Vovk and
Wang 2021; Ramdas et al. 2023).

\subsection{Finite-sample validity}\label{finite-sample-validity}

Markov's inequality is an \emph{exact, finite-sample} guarantee. The
bound \(\mathbb{P}_{H_0}(E \ge 1/\alpha) \le \alpha\) holds for any
sample size and any distribution within the null, without asymptotic
approximation. In \emph{sequential} settings, the corresponding
time-uniform guarantee is provided by Ville's inequality for e-processes
(Section Section~\ref{sec-eprocesses}): if \((E_t)\) is a nonnegative
supermartingale under \(H_0\), then
\(\mathbb{P}_{H_0}(\sup_t E_t \ge 1/\alpha) \le \alpha\).

For clinical trials, the practical consequences of this distinction
arise in three settings. First, \textbf{small samples and rare
diseases}: pediatric trials, orphan indications, and cell and gene
therapy studies often enroll tens, not hundreds, of patients per arm. At
these sample sizes, the normal approximation to a binomial test
statistic can be poor, and group sequential boundaries derived from the
asymptotic joint distribution of the test statistics may not maintain
their nominal properties (Jennison and Turnbull 2000). The e-value
guarantee holds without modification. Second, \textbf{early interim
analyses}: even in trials with large planned enrollment, the first
interim look may occur after relatively few patients. At information
fraction 0.1 of a 200-per-arm trial, the look occurs at \(n = 20\)---a
regime where asymptotic normality is suspect, particularly for binary
endpoints with response rates near 0 or 1. E-value monitoring provides
exact Type I error control at this look without relying on
distributional approximations. Third, \textbf{adaptive designs with
complex data dependencies}: response-adaptive randomization,
biomarker-driven enrichment, and platform trials with shared controls
create correlation structures that complicate the asymptotic arguments
underlying standard sequential methods. The e-value's finite-sample
guarantee is invariant to these complexities, provided the e-process
remains a supermartingale under the null.

Of course, \(p\)-values \emph{can} also be exactly valid in finite
samples for certain tests---permutation tests and exact binomial tests
are classical examples. But achieving exact validity with \(p\)-values
requires choosing from a restricted class of tests and often sacrifices
power or flexibility. E-values obtain finite-sample validity ``for
free'' from their definition: any nonnegative statistic with expected
value at most 1 under the null is immediately a valid e-value. In the
\emph{sequential} setting, the distinction is sharper: e-values, via
Ville's inequality for nonnegative supermartingales, give anytime-valid
guarantees at every stopping time simultaneously, while \(p\)-values
generally require either a fixed sample size or corrections such as
alpha-spending that must be pre-specified (Lan and DeMets 1983; Jennison
and Turnbull 2000).

\subsection{Connecting to concepts you already
use}\label{connecting-to-concepts-you-already-use}

Clinical trial biostatisticians are fluent in group sequential
monitoring, alpha-spending, conditional power, and Bayesian posterior
probabilities. E-values map naturally onto each of these:

\begin{longtable}[]{@{}
  >{\raggedright\arraybackslash}p{(\linewidth - 2\tabcolsep) * \real{0.5000}}
  >{\raggedright\arraybackslash}p{(\linewidth - 2\tabcolsep) * \real{0.5000}}@{}}
\toprule\noalign{}
\begin{minipage}[b]{\linewidth}\raggedright
Concept you know
\end{minipage} & \begin{minipage}[b]{\linewidth}\raggedright
E-value counterpart
\end{minipage} \\
\midrule\noalign{}
\endhead
\bottomrule\noalign{}
\endlastfoot
Test statistic (e.g., \(Z\)-score) at an interim & E-value \(E_n\) at
patient \(n\) \\
Rejection boundary (e.g., O'Brien--Fleming \(Z_k\)) & Fixed threshold
\(1/\alpha\) (e.g., 40 for one-sided \(\alpha = 0.025\)) \\
Alpha-spending function \(\alpha^*(t)\) & \emph{Not needed}---the
threshold is constant across all looks \\
Pre-specified number of interim looks & \emph{Not needed}---look as
often as desired \\
Design alternative (e.g., \(\delta = 0.15\)) & Design alternative used
to choose the betting fraction \(\lambda\) \\
Expected sample size under \(H_1\) &
\(N \approx \log(1/\alpha)/g(\lambda)\), where \(g(\lambda)\) is the
expected log-growth per patient pair \\
Conditional power at an interim & The current e-value indicates
accumulated evidence; \(g(\lambda)\) indicates the rate of future
evidence growth \\
Posterior probability \(\Pr(\delta > 0 \mid \text{data})\) & E-values
are frequentist, but connect to Bayes factors (see below) \\
\(p\)-value at a fixed analysis & \(1/E\) is always a valid \(p\)-value,
with anytime validity under sequential monitoring (see below) \\
\end{longtable}

The key simplification is that the \textbf{rejection boundary does not
change} from look to look. In a group sequential design, you budget your
Type I error across pre-planned interims using an alpha-spending
function: early looks use conservative boundaries (high \(Z\)-scores),
and later looks become more liberal. With e-values, the threshold stays
at \(1/\alpha\) regardless of when or how often you look. The
mathematical machinery that makes this possible---the supermartingale
property---handles the ``spending'' automatically.

\subsection{How it works: the betting interpretation}\label{sec-betting}

The most intuitive construction of an e-value for a clinical trial uses
the \textbf{testing by betting} framework (Shafer 2021; Shafer and Vovk
2019). Imagine you start with \$1 and repeatedly bet against the null
hypothesis as patients are enrolled:

\begin{itemize}
\tightlist
\item
  \textbf{After each patient pair}, you observe whether treatment or
  control did better, and you wager a fraction \(\lambda\) of your
  current wealth that treatment is superior.
\item
  \textbf{If the null is true} (no treatment effect), each bet is fair:
  on average, you neither gain nor lose money.
\item
  \textbf{If treatment is truly better}, your bets tend to pay off, and
  your wealth grows over time.
\item
  \textbf{Your current wealth is the e-value.} When it reaches \$40
  (i.e., \(1/\alpha\) for \(\alpha = 0.025\)), you have accumulated
  enough evidence to reject the null.
\end{itemize}

For exposition, we describe the trial as producing \emph{patient pairs}
(one outcome per arm) so that each step compares treatment and control
directly. This pairing is not essential: e-processes can be updated
whenever new outcome information arrives (including when accrual is
unbalanced across arms), provided the update is constructed to remain a
nonnegative supermartingale under the null and the betting choice for
the next update is predictable (measurable with respect to the past).

Formally, for a trial with 1:1 randomization and binary outcomes, define
\(D_i = X_i^T - X_i^C\) for the \(i\)-th patient pair. Under
\(H_0: p_T = p_C\), we have \(\mathbb{E}[D_i] = 0\) regardless of the
common response rate. The bettor's wealth after \(n\) patient pairs
(i.e., \(n\) outcomes per arm) is

\[
E_n = \prod_{i=1}^n (1 + \lambda_i D_i),
\]

where \(\lambda_i \in (0, 1)\) is the betting fraction chosen
\emph{before} seeing \(D_i\). The endpoints are excluded:
\(\lambda_i = 0\) produces no evidence, and \(\lambda_i = 1\) risks
zeroing out the wealth permanently if \(D_i = -1\). Because
\(D_i \in \{-1,0,1\}\) and \(\lambda_i < 1\), each factor
\((1 + \lambda_i D_i)\) is strictly positive. Under \(H_0\), we have
\(\mathbb{E}[D_i \mid D_1,\ldots,D_{i-1}] = 0\), so \((E_n)\) is a
nonnegative martingale (and hence an e-process; defined precisely
below). Two crucial points for the clinical reader:

\begin{enumerate}
\def\labelenumi{\arabic{enumi}.}
\item
  \textbf{The null is composite}---the common response rate
  \(p_T = p_C = p\) is a nuisance parameter. The betting construction
  handles this automatically because \(\mathbb{E}[D_i] = 0\) for
  \emph{every} value of \(p\), without conditioning, stratification, or
  estimation. This is analogous to how a permutation test is valid
  without knowing the underlying distribution, but here the validity
  extends to sequential monitoring.
\item
  \textbf{The betting fraction \(\lambda\) plays the role of the design
  alternative.} A biostatistician designing a group sequential trial
  must specify an expected effect size (say,
  \(\delta = p_T - p_C = 0.15\)) for the power calculation. Similarly,
  choosing \(\lambda\) encodes how much evidence you expect per patient
  pair. The ``right'' \(\lambda\) depends on the anticipated effect
  size, just as the ``right'' sample size does (see ``Choosing the
  betting fraction'' below).
\end{enumerate}

\subsection{A concrete example}\label{a-concrete-example}

Suppose we design a trial for a new oncology treatment where we expect
response rates of \(p_T = 0.35\) (treatment) vs.~\(p_C = 0.20\)
(control). In a standard group sequential design with O'Brien--Fleming
boundaries and three planned interims, we would typically choose a
maximum sample size to achieve a target power (e.g., 80\%) at
\(\alpha = 0.025\) (one-sided).

With e-value monitoring, we choose the GROW-optimal betting fraction
\(\lambda^* \approx 0.37\) (derived below) and track the wealth process
\(E_n\) after each patient pair. Consider one possible trajectory
(illustrative numbers):

\begin{itemize}
\tightlist
\item
  After 50 pairs: \(E_{50} = 3.1\) (below 40; continue)
\item
  After 100 pairs: \(E_{100} = 11.7\) (growing, but below 40; continue)
\item
  After 120 pairs: \(E_{120} = 45.2\) (exceeds \(1/0.025 = 40\); reject
  \(H_0\))
\end{itemize}

The trial could stop at patient pair 120---or at any other point where
the e-value crosses the threshold. If the DSMB also wants to look at
pairs 75 and 90 (unplanned), this remains valid for the e-process
decision rule without additional correction. The expected number of
patient pairs to rejection under the design alternative is approximately
\(\log(40)/g(\lambda^*) \approx 131\) pairs, while retaining unlimited
monitoring flexibility.

\subsection{The e-value as a sequential test statistic:
e-processes}\label{sec-eprocesses}

An \textbf{e-process} is simply a sequence of e-values
\((E_1, E_2, \ldots)\) computed as data accumulates, with the property
that stopping at \emph{any} time---planned, unplanned, or
data-dependent---yields a valid e-value. The wealth process
\(E_n = \prod_{i=1}^n (1 + \lambda_i D_i)\) from the betting
construction is an e-process. The mathematical guarantee is Ville's
inequality (Ville 1939):

\[
\mathbb{P}_{H_0}\!\left(\text{$E_n \ge 1/\alpha$ at any time $n$}\right) \le \alpha.
\]

This is the e-value analogue of the alpha-spending guarantee in group
sequential methods, but it is \emph{stronger}: it holds for every
possible stopping rule simultaneously, not just for a pre-specified set
of interim looks. The cost of this universality is that, when a fixed
look schedule is known in advance, a group sequential boundary can be
more powerful than a single time-uniform e-value threshold; our
numerical study quantifies this tradeoff.

For clinical readers familiar with Bayesian monitoring: an e-process is
a \emph{frequentist} object that provides Type I error control without
simulation-based calibration. Bayesian posterior probability thresholds
(e.g., \(\Pr(\delta > 0 \mid \text{data}) > 0.95\)) do not automatically
control Type I error under arbitrary stopping---they must be calibrated
by extensive simulation for each specific design (D. A. Berry 2025; U.S.
Food and Drug Administration 2026). The e-process gives this guarantee
by construction.

\subsection{Choosing the betting fraction: the role of the design
alternative}\label{choosing-the-betting-fraction-the-role-of-the-design-alternative}

In group sequential design, the trial statistician specifies a design
alternative (the clinically meaningful effect size) to compute the
required sample size for a target power. The analogous step in e-value
monitoring is choosing the \textbf{betting fraction} \(\lambda\), which
determines how aggressively you bet against the null.

Under the alternative \(p_T > p_C\), for \(\lambda \in (0,1)\), each
patient pair contributes an expected log-evidence of

\[
g(\lambda) = p_T(1{-}p_C)\log(1{+}\lambda) + (1{-}p_T)p_C\log(1{-}\lambda),
\]

where the two remaining outcomes---both respond (probability
\(p_T p_C\)) and neither responds (probability
\((1-p_T)(1-p_C)\))---yield \(D_i = 0\) and hence contribute
\(\log(1+\lambda \cdot 0) = 0\) to the expected growth rate. This is the
\textbf{growth rate}: the average amount of evidence (in log-units)
gained per patient pair (one observation per arm) under the alternative.
The growth-rate-optimal (GROW) betting fraction \(\lambda^*\) maximizes
\(g(\lambda)\):

\[
\lambda^* = \frac{p_T(1{-}p_C) - (1{-}p_T)p_C}{p_T(1{-}p_C) + (1{-}p_T)p_C}.
\]

For our example (\(p_T = 0.35\), \(p_C = 0.20\)), we get
\(a = 0.35 \times 0.80 = 0.28\), \(b = 0.65 \times 0.20 = 0.13\), so
\(\lambda^* = 0.15/0.41 \approx 0.37\), and the growth rate is
\(g(\lambda^*) \approx 0.028\) nats per patient pair. The expected
number of patient pairs to rejection is then

\[
N_{\text{expected}} \approx \frac{\log(1/\alpha)}{g(\lambda^*)} = \frac{\log(40)}{0.028} \approx 131,
\]

which is the e-value analogue of the expected sample size in a power
calculation. The relationship \(N \approx \log(1/\alpha)/g(\lambda)\) is
the fundamental design equation: a larger effect size yields a larger
growth rate and hence a shorter trial, exactly as in classical design.

If the true effect is larger than anticipated, the e-value grows faster
and the trial stops sooner. If the true effect is smaller, the e-value
grows slowly---and this is where the betting fraction matters. Choosing
\(\lambda\) too large (over-betting) is like powering a trial for a
large effect that doesn't exist: the e-value oscillates and may never
cross the threshold. Choosing \(\lambda\) too small (under-betting) is
like running an overpowered trial: you will eventually reject, but waste
patients. In practice, one can start with a design-alternative-based
\(\lambda\) and adaptively adjust it using accumulating data, provided
the adaptation rule is predictable (i.e., uses only information
available before observing the next outcome).

\subsection{Relationship to p-values and Bayes
factors}\label{relationship-to-p-values-and-bayes-factors}

The reciprocal \(p' = 1/E\) of any e-value is a valid \(p\)-value in the
usual fixed-horizon sense: for any \emph{pre-specified}
\(\alpha \in (0,1)\),
\(\mathbb{P}_{H_0}(p' \le \alpha) = \mathbb{P}_{H_0}(E \ge 1/\alpha) \le \alpha\).
In sequential monitoring, an always-valid \(p\)-value process is
obtained by \(p_t = 1/\sup_{s \le t} E_s\), which controls Type I error
under optional stopping (Vovk and Wang 2021; Shafer 2021). As with all
\(p\)-values, however, choosing the rejection level \(\alpha\)
\emph{after} seeing the data is not generally a valid inferential
operation; in practice, e-values are best treated as the primary
evidential scale and converted to \(p\)-values as a reporting layer when
needed. Calibrator functions can also convert standard \(p\)-values into
e-values, though the resulting e-values are typically less powerful than
purpose-built ones (Vovk and Wang 2021).

E-values also connect to \textbf{Bayes factors}. When the null
hypothesis is simple (a single distribution), any Bayes factor is a
valid e-value. When the null is composite---as in most clinical
trials---the relationship is more subtle: most Bayes factors are
\emph{not} e-values, and most e-values are not Bayes factors (Ramdas et
al. 2023; Polson, Sokolov, and Zantedeschi 2026; Datta et al. 2025). The
method of mixtures construction (used in the safe logrank test for
survival endpoints; Grünwald et al. (2021)) produces e-values that
resemble Bayes factors---they integrate the likelihood over a mixing
distribution on the alternative---but the mixing distribution is chosen
to maximize evidence growth rather than to reflect prior beliefs. For
Bayesian trialists, this is a familiar mathematical form with a
frequentist validity guarantee.

\subsection{Three ways to build
e-values}\label{three-ways-to-build-e-values}

Beyond the betting construction (which we use for binary endpoints), two
other approaches are useful in clinical settings (Ramdas et al. 2023;
Grünwald, Heide, and Koolen 2023):

\textbf{Method of mixtures.} Specify a ``mixing'' distribution \(W\)
over plausible treatment effects and average the likelihood ratio
against this distribution: \(E = \int L(\theta)\, dW(\theta)\), where
\(L(\theta)\) is the likelihood ratio for a specific effect size
\(\theta\). This resembles computing a Bayes factor, but the mixing
distribution is chosen for statistical efficiency, not for representing
prior beliefs. The \textbf{safe logrank test} for time-to-event
endpoints uses this approach, with a mixing distribution centered on the
anticipated hazard ratio (Grünwald et al. 2021). For trial statisticians
comfortable with Bayesian analyses, this construction is the most
familiar pathway to e-values.

\textbf{Calibrators (p-to-e conversion).} If you already have a valid
\(p\)-value \(P\) from a standard analysis, you can obtain an e-value by
applying a \emph{calibrator} function \(f:[0,1]\to[0,\infty)\) that is
nonincreasing and satisfies \(\int_0^1 f(u)\,du \le 1\), and setting
\(E = f(P)\). This provides a simple way to wrap an existing test in an
e-value layer, at some cost in efficiency relative to purpose-built
e-values derived from likelihood or betting constructions (Vovk and Wang
2021; Ramdas et al. 2023).

Bayes factors are likelihood ratios of marginal likelihoods under two
models and can serve as e-values when formulated against a simple null
or when they satisfy the e-value expectation bound. Bayesian decision
rules used in adaptive trials, however, often rely on posterior or
predictive probabilities crossing fixed thresholds (e.g.,
\(\mathbb{P}(p_T > p_C \mid \text{data}) > 0.95\)). Such thresholds do
not automatically yield frequentist error control under optional
stopping without calibration, typically obtained by extensive simulation
(D. A. Berry 2025; U.S. Food and Drug Administration 2025, 2026). In
contrast, e-values deliver a direct, analytic Type I error guarantee via
Ville's inequality, but may require careful tuning to achieve
competitive power.

\section{E-values for adaptive clinical
trials}\label{e-values-for-adaptive-clinical-trials}

\subsection{Interim monitoring, protocol adaptations, and
response-adaptive
randomization}\label{interim-monitoring-protocol-adaptations-and-response-adaptive-randomization}

In many adaptive trials, the number and timing of interim analyses are
not strictly fixed (e.g., due to information-driven DSMB scheduling,
delayed outcomes, or operational constraints). E-process monitoring
permits arbitrary additional looks without re-deriving spending
boundaries, provided the e-process remains valid under the filtration
that includes the adaptation decisions. The threshold
\(E_t \ge 1/\alpha\) does not change regardless of whether two or twenty
interim looks are conducted, and no alpha-spending function needs to be
specified or recalibrated. This permits continuous monitoring for safety
or efficacy under a single threshold.

Adaptations such as sample size modification, arm dropping, enrichment,
or response-adaptive randomization (RAR) are typically functions of past
information. In martingale language, these are predictable design
choices. E-process validity is preserved under such predictable
adaptations as long as the e-process remains a supermartingale under
\(H_0\) relative to the enlarged filtration.

Response-adaptive randomization (RAR) presents a particular challenge
for both classical and e-value-based methods. Berry traces the
intellectual lineage of RAR to Thompson's 1933 proposal (Thompson 1933;
D. A. Berry 2025), which assigned the next patient to the arm with
probability equal to the current Bayesian probability that the arm is
superior---a strategy that, in Berry's words, ``predated Hill's
publication by 15 years'' and is ``more flexible and arguably more
useful than Hill's'' fixed randomization. Thompson sampling has been
deployed at scale in modern platform trials: in I-SPY 2, adaptive
randomization directed more patients to promising arms within their
molecular subtypes, and Berry reports that for neratinib, 10 of the 17
percentage-point improvement in pathological complete response rate over
control was attributable to adaptive randomization rather than the
drug's intrinsic effect (Barker et al. 2009; Park et al. 2016; D. A.
Berry 2025). For e-processes under RAR, the key requirement is that the
betting fraction \(\lambda_i\) must be predictable---determined before
observing \(D_i\)---but it may depend on the allocation probabilities
used for patient \(i\), which are themselves functions of past data. In
the simplest case (fixed \(\lambda\)), this is automatically satisfied.
With adaptive \(\lambda_i\), care is needed to ensure that the
adaptation rule for \(\lambda_i\) and the RAR rule are both predictable
with respect to the same filtration. Thompson sampling and the
testing-by-betting tradition share a common structure: both involve
making decisions (treatment allocation or evidence accumulation) by
computing quantities that depend on the current posterior, both are
inherently sequential, and both satisfy predictability by construction.
Recent work on exact frequentist analysis for response-adaptive designs
(Baas, Jacko, and Villar 2025) complements the e-value approach by
providing computationally tractable exact tests that can be used as
benchmarks for calibration.

\subsection{Time-to-event endpoints and safe logrank
testing}\label{sec-safe-logrank}

For survival endpoints, sequential monitoring is standard, but the safe
logrank test provides an e-value-based analogue with Type I error
guarantees under optional stopping and continuation (Grünwald et al.
2021). The construction proceeds as follows. At each observed event
time, the logrank contribution is the difference between the observed
and expected number of events in the treatment arm, conditional on the
risk sets. Under the null of equal hazards, this contribution has mean
zero, providing the basis for a betting-type e-process. The safe logrank
test uses a mixture likelihood ratio, integrating over a prior on the
log-hazard ratio, to obtain an e-value that can be computed
incrementally as events accumulate. The \texttt{safestats} R package
(Schure and Ly 2022) provides an implementation with tools for
specifying the design alternative (as a hazard ratio and prior
concentration), computing expected stopping times, and calibrating the
prior for target power.

Compared with the standard group sequential logrank test, the safe
logrank test sacrifices some power at any fixed analysis time but gains
the ability to continue data collection beyond the planned maximum
without invalidating inference---a property that is operationally
valuable when event accrual is slower than anticipated or when external
circumstances delay the planned analysis. A practical issue that arises
in survival trials is delayed outcomes: when the primary endpoint is
overall survival, many patients at each interim analysis have been
treated but have not yet experienced an event. Berry's approach to this
problem, used in ASTIN, AWARD-5, GBM AGILE, and the Leqembi 201
Alzheimer's trial, is to build longitudinal models that predict late
endpoints from early measurements and use Bayesian multiple imputation
to incorporate partial follow-up data into interim decisions (Geiger et
al. 2012; Alexander et al. 2018; D. A. Berry et al. 2023; D. A. Berry
2025). For e-process methodology, delayed outcomes create a filtration
gap: the e-process at time \(t\) should ideally incorporate all
information available at \(t\), but some outcome data is missing not at
random. One direction for integrating e-values with Bayesian
longitudinal modeling is to construct the betting increment at each
event time using the best available prediction of the outcome, with the
e-process remaining valid under the null as long as the imputation model
is predictable.

\subsection{Futility monitoring via
e-processes}\label{futility-monitoring-via-e-processes}

Futility stopping is complementary to efficacy stopping and is equally
important for ethical and economic reasons. E-processes support futility
through two routes.

The first route uses confidence sequences. At each interim look, one
inverts the e-process to obtain a \((1-\alpha)\) confidence sequence
\(C_n\) for the treatment effect \(\delta = p_T - p_C\) (see Section
Section~\ref{sec-reporting}). If the upper bound of \(C_n\) falls below
the minimum clinically important difference \(\delta_{\min}\), the trial
can be stopped for futility with the assurance that, at the current
evidence level, a clinically meaningful effect is incompatible with the
data. Because the confidence sequence is anytime-valid, this futility
rule does not require pre-specification of the number or timing of
futility looks.

The second route constructs a separate e-process for the ``reverse''
null \[
H_0'{:}\, p_T - p_C \ge \delta_{\min}.
\] Define the centered increment \[
D_i' = \delta_{\min} - (X_i^T - X_i^C).
\] Under \(H_0'\), we have \[
\mathbb{E}[D_i'] = \delta_{\min} - (p_T - p_C) \le 0,
\] so \(D_i'\) has non-positive mean. A betting martingale
\(E_n' = \prod_{i=1}^n (1 + \lambda' D_i')\) with
\(\lambda' \in (0,\, 1/(1 - \delta_{\min})]\) (this upper bound ensures
\(1 + \lambda' D_i' \ge 0\) for all
\(D_i' \in [\delta_{\min} - 1,\, \delta_{\min} + 1]\)) is a nonnegative
supermartingale under \(H_0'\). Note that \(D_i'\) depends only on
\(\delta_{\min}\) (a prespecified design parameter) and the observed
data, not on any unknown nuisance parameters. When the true effect is
smaller than \(\delta_{\min}\), \(\mathbb{E}[D_i'] > 0\) and the wealth
process grows, accumulating evidence for futility. If \(E_n'\) exceeds
\(1/\alpha_f\) (where \(\alpha_f\) is the futility error budget), there
is strong evidence that the treatment does not achieve a clinically
meaningful effect.

\subsection{Multiplicity and platform
trials}\label{multiplicity-and-platform-trials}

Platform trials monitor multiple hypotheses across arms, subpopulations,
and endpoints. E-values admit combination operations that are
algebraically simpler than their \(p\)-value analogues: the arithmetic
mean of e-values for the same null remains a valid e-value under
arbitrary dependence (Vovk and Wang 2021; Ramdas et al. 2023), and the
product of independent e-values remains a valid e-value. This supports
false discovery rate control via procedures designed for e-values (e.g.,
e-BH), with formal guarantees under the conditions described in Wang and
Ramdas (2022).

To illustrate concretely, consider a platform trial with a shared
control arm and \(K\) experimental arms---the architecture used in I-SPY
2, GBM AGILE, REMAP-CAP/COVID, and Precision Promise (Barker et al.
2009; Park et al. 2016; Alexander et al. 2018; Angus et al. 2020; D. A.
Berry 2025). Woodcock and LaVange (Woodcock and LaVange 2017) formalized
the taxonomy of such master protocols and used I-SPY 2 as a prototypic
example. For each arm \(k\), a betting e-process \(E_n^{(k)}\) is
maintained against \(H_0^{(k)}{:}\, p_k = p_C\). Arms are graduated when
\(E_n^{(k)} \ge 1/\alpha_k\) and dropped when a futility criterion is
met. When an arm is dropped, its e-process is frozen; when a new arm
enters, a new e-process is initialized. Because each e-process provides
an anytime-valid level-\(\alpha_k\) test for its own hypothesis, the
familywise error rate across graduated arms is controlled at
\(\sum_k \alpha_k\) by a union bound (with \(\alpha_k\) allocated across
arms). Shared controls induce dependence between hypotheses; if the
design goal is false discovery rate control across many arms or
subgroups, e-values can be filtered using e-value FDR procedures (e.g.,
e-BH), with guarantees under the conditions described in Wang and Ramdas
(2022). The \texttt{evalinger} functions \texttt{platform\_monitor()}
and \texttt{ebh()} implement multi-arm e-process monitoring with e-BH
multiplicity control for this setting. Platform trials may also borrow
information from nonconcurrent controls or other sources via
hierarchical modeling and time-trend adjustment (Saville et al. 2022; D.
A. Berry 2025). Such borrowing changes the dependence structure of the
monitoring problem; e-process validity then requires that the borrowing
mechanism and any adaptive weighting be predictable with respect to the
monitoring filtration. The draft FDA Bayesian guidance discusses
nonconcurrent controls and borrowing considerations in platform
settings, including effective sample size accounting and sensitivity
analyses for prior-data conflict (U.S. Food and Drug Administration
2026; Morita, Thall, and Müller 2008; Evans and Moshonov 2006). In such
settings, an e-process can serve as an additional, time-uniform
evidential check alongside the Bayesian operating-characteristics
analyses used to design and justify the borrowing model.

\section{Reporting, estimation, and calibration}\label{sec-reporting}

Clinical trial reporting is organized around familiar objects:
\(p\)-values, confidence intervals, and prespecified decision
thresholds. When an e-process is used for monitoring, an always-valid
\(p\)-value process is obtained by setting
\(p_t = 1/\sup_{s \le t} E_s\), which satisfies
\(\mathbb{P}_{H_0}(\inf_t p_t \le \alpha) \le \alpha\) by Ville's
inequality (Vovk and Wang 2021; Shafer 2021). These conversions are
useful when the trial's communication context demands \(p\)-values, but
they should be viewed as a reporting layer on top of the underlying
e-process rather than as the primary design object.

Sequential monitoring is rarely only about testing; it is also about
estimating treatment effects with quantifiable uncertainty at each
interim look. E-processes yield always-valid confidence sequences
through test inversion. For a parameter \(\theta\) indexing the
treatment effect, one constructs an e-process \(E_t(\theta)\) for each
hypothesized value and defines
\(C_t = \{\theta : \sup_{s \le t} E_s(\theta) < 1/\alpha\}\). Under
regularity conditions, \(C_t\) is a \((1-\alpha)\) confidence sequence:
the probability that the true parameter ever leaves \(C_t\) is at most
\(\alpha\) (Howard et al. 2021; Ramdas et al. 2023). Such sequences
provide a time-uniform analogue of confidence intervals and can be
reported alongside monitoring decisions to contextualize effect size
estimates at interim analyses.

The practical advantages of e-values do not remove the need for design
calibration; they change its target. The Type I error guarantee is
analytic and does not depend on a spending schedule, but power and
expected sample size depend on how the e-process is constructed. In
parametric settings this dependence takes the form of a choice of
alternative model or betting strategy; in nonparametric settings it may
take the form of a regularization scheme. Consequently, the design task
shifts from selecting an alpha-spending boundary to selecting and tuning
an e-process, with performance assessed via simulation in clinically
plausible scenarios (Grünwald, Heide, and Koolen 2023; Martin 2025).

\section{Optional continuation and evidence
synthesis}\label{optional-continuation-and-evidence-synthesis}

An important distinction between e-values and classical test outputs is
that e-values can be meaningfully accumulated across an adaptively
determined sequence of experiments. If each component yields a valid
e-value \(E^{(k)}\) for the same null claim, then under standard
independence or conditional independence conditions, the product
\(E^{(1)} E^{(2)} \cdots E^{(K)}\) remains an e-value even when \(K\) is
data-dependent and the design of study \(k+1\) is chosen after observing
earlier outcomes (Grünwald, Heide, and Koolen 2023; Ramdas et al. 2023).
This property provides a formal route to sequential meta-analytic
thinking: evidence can be updated as data accrue over time and across
study boundaries without invalidating Type I error control.

In clinical development, product accumulation should be viewed as a
principled evidential scaffold rather than as an automatic license for
pooling, because the null claim is rarely literally identical across
studies when populations, endpoints, or operational definitions change.
Practical use requires careful articulation of the claim being tested,
including how the estimand is defined across components and what forms
of variation are tolerated. Nonetheless, the optional continuation
property offers a mathematically coherent way to track evidence over
time in settings where fixed-horizon analyses encourage informal and
error-prone evidence aggregation.

\section{Implementation, software, and regulatory
context}\label{sec-regulatory}

Because the e-process guarantee is time-uniform, the operational
monitoring plan can be simplified: interim looks may occur at arbitrary
times, but the monitoring statistic and threshold do not change. A
monitoring charter should specify the filtration explicitly, including
which data streams and covariates are incorporated at each interim
analysis, how delayed outcomes are handled, and which adaptation
decisions are permitted as predictable functions of past information.
Computation should be separated from decision-making authority:
e-processes are computed from the accumulating data, but decisions to
stop, continue, or modify the trial remain governance actions that must
be justified clinically and ethically. Reporting \(\log E_t\) at each
interim look yields an additive evidence trajectory that is easier to
interpret than a sequence of binary boundary crossings, and when
combined with confidence sequences, this simultaneously conveys
evidential strength and uncertainty about effect size.

Software for e-value construction in clinical trials is maturing. The
\texttt{safestats} R package (Schure and Ly 2022) provides safe tests
for two-group comparisons (proportions, means, and survival) with tools
for design calibration (expected stopping time, required sample size,
and power), including the safe logrank test, safe \(z\)-tests, and safe
\(t\)-tests, and can compute optimal GROW betting fractions for
specified design alternatives. Research code accompanying Ramdas et al.
(2023) provides additional e-process constructions. Clinical trial
simulation platforms (e.g., \texttt{rpact}, \texttt{gsDesign}) do not
yet natively support e-value monitoring, but e-process computations can
be embedded in custom simulation frameworks.

To bridge the gap between the e-value literature and operational
clinical trial practice, we provide the \texttt{evalinger} R package
(source code and documentation at
\url{https://github.com/VadimSokolov/evalinger}; installable via
\texttt{devtools::install\_github("VadimSokolov/evalinger")}), which
implements the complete methodology described in this paper. The package
provides: (i) \texttt{eprocess\_binary()} and
\texttt{eprocess\_logrank()} for constructing e-processes for binary and
survival endpoints; (ii) \texttt{grow\_lambda()},
\texttt{expected\_growth\_rate()}, and \texttt{edesign\_binary()} for
GROW-optimal design calibration; (iii) \texttt{emonitor()} with
streaming \texttt{update()} for real-time sequential monitoring; (iv)
\texttt{confseq\_binary()} for time-uniform confidence sequences; (v)
\texttt{futility\_cs()} and \texttt{futility\_eprocess()} for futility
monitoring via both confidence-sequence and reciprocal-e-process routes;
(vi) \texttt{platform\_monitor()} and \texttt{ebh()} for multi-arm
platform trial monitoring with e-Benjamini--Hochberg multiplicity
control; (vii) \texttt{hybrid\_monitor()} for joint e-process and group
sequential monitoring; and (viii) \texttt{simulate\_comparison()} for
the five-method Monte Carlo comparison reported in the Numerical
Demonstration section. The package integrates with the established R
clinical trials ecosystem (\texttt{gsDesign}, \texttt{rpact},
\texttt{survival}, \texttt{safestats}). An interactive Shiny web
application, bundled with the package and hosted at
\url{https://sailtargets.shinyapps.io/evalinger/}, provides three
dashboards---Design Calculator, Monitoring Dashboard, and Method
Comparison---that allow practitioners to explore the methodology without
writing code: trialists can calibrate the GROW-optimal betting fraction
for their design alternative, simulate batch-by-batch monitoring with
live e-process trajectory and confidence sequence visualization, and run
head-to-head comparisons of e-value, group sequential, and Bayesian
monitoring rules under user-specified scenarios. The web application
supports the operating-characteristics documentation that the 2026 FDA
draft Bayesian guidance requires for Bayesian and adaptive designs (U.S.
Food and Drug Administration 2026).

From a regulatory standpoint, the FDA's guidance on adaptive designs for
drugs and biologics (U.S. Food and Drug Administration 2019) requires
that adaptive methods control Type I error in the strong sense and
maintain trial integrity. E-values satisfy the error control requirement
by construction (Ville's inequality), but the guidance does not
specifically reference e-values, and regulatory reviewers may be
unfamiliar with the methodology. Berry's experience over three decades
of introducing Bayesian methods into regulatory practice is a relevant
precedent (D. A. Berry 2025). The path from early Bayesian premarket
approvals for medical devices (including a spinal implant trial around
2000--2001, with subsequent CDRH review experience documented by
Pennello and Thompson (2007) and Irony and Campbell (2011)) through the
CDRH Bayesian guidance for devices {[}originally issued in 2010; U.S.
Food and Drug Administration (2025); Irony et al. (2023){]} to CDER's
acceptance of fully Bayesian registration trials (Pravigard Pac in 2003,
Trulicity/AWARD-5 in 2014 (Geiger et al. 2012), the Leqembi Alzheimer's
disease program (D. A. Berry et al. 2023), and GBM AGILE as a
seamless-phase registration platform (Alexander et al. 2018)) was
neither smooth nor fast, and Berry documents the need for regulatory
champions, extensive simulation-based operating characteristics, and
dual reporting in both Bayesian and frequentist terms. Berry reports
that after pushback from JAMA editors who demanded frequentist reporting
for the ThermoCool AF trial, he now caveats any required frequentist
measures with the note that ``All \(p\)-values and confidence intervals
in this article are descriptive and have no inferential content'' (D. A.
Berry 2025).

A landmark development in this trajectory is the January 2026 FDA draft
guidance on the use of Bayesian methodology in clinical trials of drugs
and biological products, issued jointly by CDER and CBER (U.S. Food and
Drug Administration 2026). This guidance formalizes decades of
regulatory experience and, as Viele notes in Berry Consultants'
commentary, represents ``a dramatic leap forward for Bayesian clinical
trials and regulatory science'' (Viele 2026). Three features of the
guidance are directly relevant to the methodology we develop here.
First, the guidance codifies three distinct pathways for success
criteria in Bayesian trials: (i) calibration to Type I error rate, where
Bayesian posterior thresholds are simulation-tuned to achieve a
one-sided 2.5\% FWER---the framework used by most Bayesian trials at FDA
to date and the approach illustrated by our calibrated Bayesian rule in
Section Section~\ref{sec-numerical}; (ii) direct interpretation of the
posterior probability, where the prior is agreed to ``provide an
accurate summary of the state of belief'' and the posterior probability
itself justifies the decision, without requiring frequentist
calibration; and (iii) benefit-risk or decision-theoretic approaches,
where success thresholds incorporate explicit loss functions balancing
the consequences of approving an ineffective drug against those of
failing to approve an effective one. E-values map most naturally onto
the first pathway, where they provide an analytic guarantee of Type I
error control that avoids the extensive simulation-based calibration the
guidance requires for posterior-threshold approaches. But they also
complement the second and third pathways: an e-process running in
parallel with a purely Bayesian decision rule provides a frequentist
evidential ledger that can reassure regulators or reviewers who are not
prepared to accept a direct posterior interpretation.

Second, the draft guidance distinguishes the \emph{analysis prior} used
in the final analysis from \emph{design priors} used to evaluate
operating characteristics under a range of plausible scenarios
(Spiegelhalter, Abrams, and Myles 2004; U.S. Food and Drug
Administration 2026). This parallels the design calibration of
e-processes: one chooses the betting strategy (or an equivalent tuning
parameter) to be efficient under a design alternative, and then
evaluates sensitivity across a range of alternatives (as in
Table~\ref{tbl-sensitivity}). The draft guidance emphasizes that
operating characteristics can be particularly sensitive to prior
assumptions in small samples and in designs with early interim analyses,
where dense or irregular monitoring can make simulation-based
calibration more fragile.

Third, the draft guidance discusses borrowing and prior-data conflict,
emphasizing design-stage exploration of operating characteristics and
sensitivity analyses via simulation (Evans and Moshonov 2006; U.S. Food
and Drug Administration 2026). This context is relevant to e-values in
borrowing settings, because e-values can be constructed conditional on
the borrowed data (treated as part of the filtration) and can provide
anytime-valid Type I error control without requiring re-derivation of
sequential thresholds when monitoring schedules change. In platform
trials where nonconcurrent control borrowing is used---including
settings referenced by the draft guidance---an e-process that accounts
for the borrowing mechanism in its filtration can provide a time-uniform
evidential guarantee that complements the Bayesian operating
characteristics documentation required by regulators (U.S. Food and Drug
Administration 2026).

E-value proponents face an analogous translation challenge to the one
Berry navigated over three decades. Practical communication strategies
include presenting the e-value monitoring plan as a pre-planned
sequential analysis method with a single time-uniform threshold,
providing always-valid \(p\)-values and confidence sequences alongside
the e-process for familiar reporting, and including simulation-based
operating characteristics (power, expected sample size, Type I error) in
the statistical analysis plan, as is standard for group sequential and
Bayesian designs. The FDA's Complex Innovative Design (CID) initiative
encourages novel trial designs for demonstrating effectiveness (U.S.
Food and Drug Administration 2019), and the 2026 draft Bayesian
guidance's openness to non-traditional success criteria may provide a
regulatory pathway for e-value-based monitoring in confirmatory trials.

\section{Numerical demonstration}\label{sec-numerical}

This section compares five monitoring rules in a two-arm randomized
controlled trial with a binary endpoint. The goal is to illustrate how
error control and power depend on the inferential object and the
stopping rule, and to quantify the operational tradeoffs among
monitoring approaches. All results are generated by the companion
scripts \texttt{run\_simulations.R} and \texttt{run\_extended\_sims.R}
using the \texttt{evalinger} package, making them fully
reproducible.\footnote{To regenerate results:
  \texttt{Rscript\ run\_simulations.R\ \&\&\ Rscript\ run\_extended\_sims.R}
  from the paper directory. Outputs are saved to \texttt{sims/}.}

\subsection{Design and monitoring
rules}\label{design-and-monitoring-rules}

We consider a two-arm RCT with \(1{:}1\) randomization to treatment (T)
and control (C) arms. Each patient's outcome is an independent Bernoulli
draw: \(X_i^T \sim \text{Ber}(p_T)\) and \(X_i^C \sim \text{Ber}(p_C)\).
We test

\[
H_0{:}\, p_T = p_C \quad \text{versus} \quad H_1{:}\, p_T > p_C,
\]

with \(p_C = 0.3\) under both hypotheses, \(p_T = 0.3\) under \(H_0\),
and \(p_T = 0.45\) under \(H_1\) (a treatment effect of 15 percentage
points). The maximum sample size is \(N_{\max} = 200\) per arm (400
total). We conduct 20 equally spaced interim analyses, one every 10
patients per arm. The nominal one-sided Type I error target is
\(\alpha = 0.025\).

The GROW-optimal betting fraction for this design alternative is
\(\lambda^* = 0.312\), with an expected growth rate of
\(g(\lambda^*) = 0.0238\) nats per patient pair and an approximate
expected stopping time of \(155\) pairs under the alternative.

The five monitoring rules are:

\begin{enumerate}
\def\labelenumi{\arabic{enumi}.}
\tightlist
\item
  \textbf{Naive repeated \(p\)-values}: at each interim look, a
  one-sided Wald \(z\)-test for \(p_T > p_C\) is computed, and the trial
  stops when the \(p\)-value falls below \(\alpha\).
\item
  \textbf{Calibrated group sequential}: the same \(z\)-statistic is
  compared to an O'Brien--Fleming-like boundary \(z \ge c/\sqrt{t}\),
  where \(t = n/N_{\max}\) is the information fraction and \(c\) is
  calibrated by Monte Carlo to achieve overall Type I error \(\alpha\).
\item
  \textbf{E-value, betting}: the betting martingale
  \(E_n = \prod_{i=1}^n (1 + \lambda D_i)\) with GROW-optimal
  \(\lambda\). Reject when \(E_n \ge 1/\alpha\).
\item
  \textbf{Naive posterior threshold}: Bayesian posterior
  \(\Pr(p_T > p_C \mid \text{data})\) with Jeffreys priors, stopping
  when it exceeds \(1 - \alpha = 0.975\).
\item
  \textbf{Calibrated Bayesian}: same posterior, but the threshold is
  simulation-calibrated to control Type I error at \(\alpha\) (S. M.
  Berry et al. 2010; U.S. Food and Drug Administration 2025, 2026).
\end{enumerate}

All five rules are implemented in the \texttt{evalinger} function
\texttt{simulate\_comparison()}.

\subsection{Results}\label{results}

Table~\ref{tbl-main} reports Monte Carlo estimates based on 50,000
simulated trials under \(H_0\) and \(H_1\). For each rule, we report the
rejection probability (Type I error under \(H_0\), power under \(H_1\))
and the average stopping sample size per arm.

\begin{longtable}[]{@{}
  >{\raggedright\arraybackslash}p{(\linewidth - 8\tabcolsep) * \real{0.4286}}
  >{\raggedleft\arraybackslash}p{(\linewidth - 8\tabcolsep) * \real{0.1190}}
  >{\raggedleft\arraybackslash}p{(\linewidth - 8\tabcolsep) * \real{0.1190}}
  >{\raggedleft\arraybackslash}p{(\linewidth - 8\tabcolsep) * \real{0.1667}}
  >{\raggedleft\arraybackslash}p{(\linewidth - 8\tabcolsep) * \real{0.1667}}@{}}

\caption{\label{tbl-main}Comparison of five monitoring rules in a
two-arm binary RCT (\(p_C = 0.30\), design alternative \(p_T = 0.45\),
\(N_{\max} = 200\) per arm, 20 looks, \(\alpha = 0.025\)). Rejection
probabilities and average sample sizes (per arm) estimated from 50,000
Monte Carlo replications.}

\tabularnewline

\toprule\noalign{}
\begin{minipage}[b]{\linewidth}\raggedright
Monitoring rule
\end{minipage} & \begin{minipage}[b]{\linewidth}\raggedleft
Null rej.
\end{minipage} & \begin{minipage}[b]{\linewidth}\raggedleft
Alt. rej.
\end{minipage} & \begin{minipage}[b]{\linewidth}\raggedleft
Avg. n (null)
\end{minipage} & \begin{minipage}[b]{\linewidth}\raggedleft
Avg. n (alt.)
\end{minipage} \\
\midrule\noalign{}
\endhead
\bottomrule\noalign{}
\endlastfoot
E-value, betting (\(\lambda = 0.31\)) & 0.012 & 0.723 & 199.013 &
139.230 \\
Calibrated group sequential (OBF) & 0.025 & 0.861 & 198.930 & 139.819 \\
Naive repeated \(p\)-values & 0.148 & 0.933 & 178.376 & 74.206 \\
Naive posterior threshold (0.975) & 0.135 & 0.932 & 181.013 & 76.233 \\
Calibrated Bayesian & 0.020 & 0.688 & 197.085 & 133.913 \\

\end{longtable}

The naive repeated \(p\)-value rule and the naive posterior-probability
rule both exhibit pronounced Type I error inflation under frequent
interim looks. The calibrated group sequential boundary controls Type I
error by design and achieves the highest power among the
error-controlling methods. The calibrated Bayesian rule also controls
Type I error (by construction via its simulation-calibrated threshold)
with the notable advantage of earlier stopping under the alternative.
The e-value rule controls Type I error with a substantial margin in this
simulation, reflecting the conservatism of this particular e-process
(Ville's inequality need not be tight).

\begin{figure}[H]

\centering{

\includegraphics[width=1\linewidth,height=\textheight,keepaspectratio]{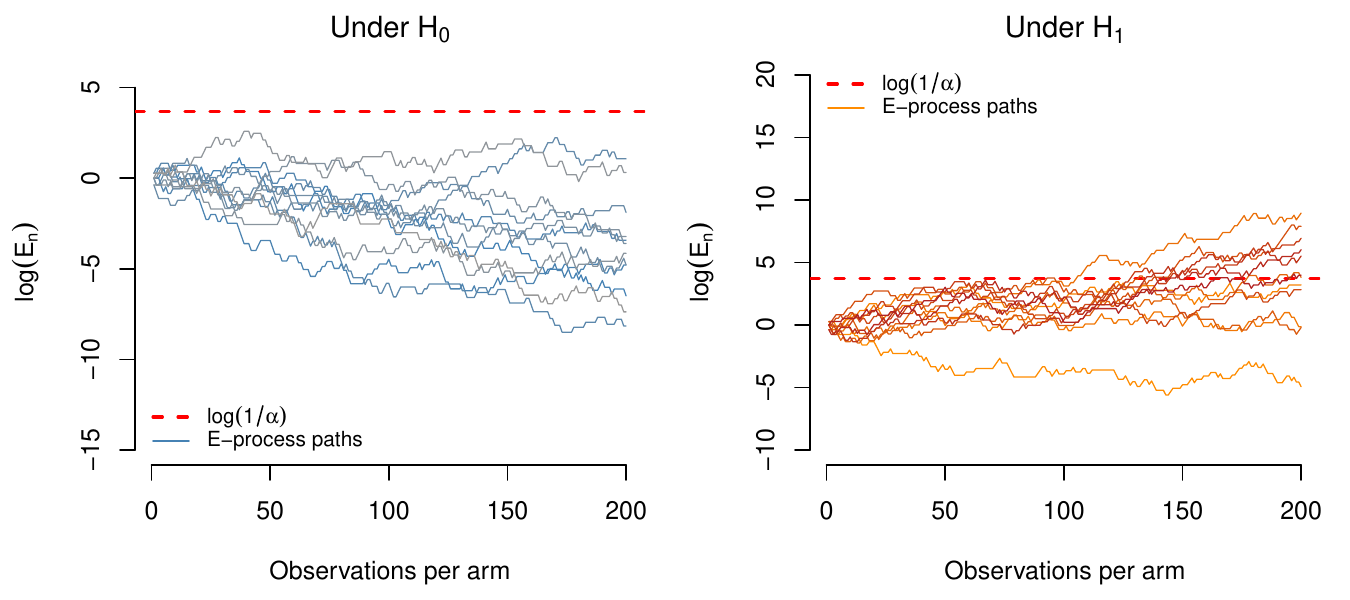}

}

\caption{\label{fig-paths}Sample paths of the log-e-process \(\log E_n\)
for 12 simulated trials under the null (left) and the alternative
(right). The horizontal dashed line marks the rejection threshold
\(\log(1/\alpha)\). Under \(H_0\), the bettor's log-wealth drifts
downward; under \(H_1\), most paths cross the threshold before the
maximum sample size.}

\end{figure}

Figure~\ref{fig-paths} displays sample paths of \(\log E_n\) for 12
simulated trials under \(H_0\) and \(H_1\). Under the null, the
log-e-process drifts downward (the bettor loses wealth on average when
the null is true). Under the alternative, paths tend upward and most
cross the threshold well before \(N_{\max}\), illustrating the
accumulation of evidence against a false null.

\subsection{Sensitivity to betting
fraction}\label{sensitivity-to-betting-fraction}

The e-value monitoring rule depends on the betting fraction \(\lambda\),
which plays a role analogous to the choice of spending function in group
sequential design or the prior in Bayesian monitoring.
Table~\ref{tbl-sensitivity} shows how power and expected sample size
vary with \(\lambda\).

\begin{longtable}[]{@{}rrrrr@{}}

\caption{\label{tbl-sensitivity}Sensitivity of e-value monitoring to the
betting fraction \(\lambda\). Type I error remains below
\(\alpha = 0.025\) for all choices, but power varies substantially. The
GROW-optimal \(\lambda\) maximizes power, while conservative or
aggressive choices lose substantial power.}

\tabularnewline

\toprule\noalign{}
\(\lambda\) & Null rej. & Alt. rej. & Avg. n (null) & Avg. n (alt.) \\
\midrule\noalign{}
\endhead
\bottomrule\noalign{}
\endlastfoot
0.10 & 0.000 & 0.119 & 200.000 & 197.495 \\
0.20 & 0.004 & 0.642 & 199.828 & 161.593 \\
0.31 & 0.012 & 0.722 & 199.048 & 139.825 \\
0.40 & 0.014 & 0.685 & 198.521 & 135.428 \\
0.50 & 0.014 & 0.589 & 198.190 & 139.088 \\

\end{longtable}

A conservative choice (\(\lambda = 0.10\)) yields almost no power,
because the per-observation growth rate is too slow to overcome the
rejection threshold in 200 observations. The GROW-optimal
\(\lambda \approx 0.31\) maximizes power. Aggressive choices
(\(\lambda \ge 0.40\)) sacrifice power because the increased variance of
the log-e-process outweighs the higher expected drift. In all cases,
Type I error remains well below \(\alpha\), confirming the robustness of
anytime-valid control to the betting fraction. This sensitivity analysis
shows that calibrating \(\lambda\) to the design alternative is
essential, in the same spirit that group sequential designs calibrate
the spending function and Bayesian designs calibrate the prior.

\begin{figure}[H]

\centering{

\includegraphics[width=0.8\linewidth,height=\textheight,keepaspectratio]{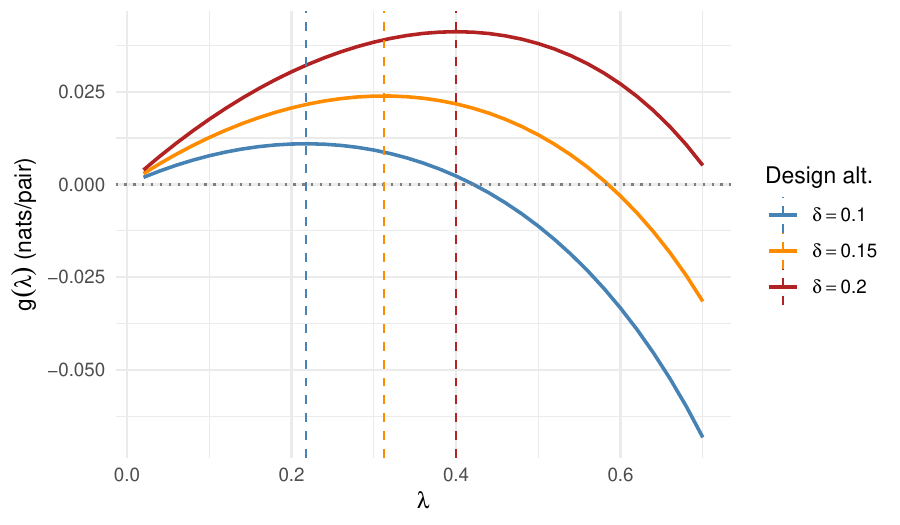}

}

\caption{\label{fig-growth-landscape}Expected log-growth rate
\(g(\lambda;\, p_T, p_C)\) as a function of the betting fraction
\(\lambda\) for three design alternatives
(\(\delta = p_T - p_C = 0.10, 0.15, 0.20\) with \(p_C = 0.30\)). The
vertical dashed lines mark the GROW-optimal \(\lambda^*\) for each
alternative.}

\end{figure}

The theoretical growth rate landscape
(Figure~\ref{fig-growth-landscape}) visualizes why the GROW-optimal
\(\lambda\) matters: over-betting (\(\lambda\) too large) can make the
growth rate negative, while under-betting wastes evidence.

\subsection{Hybrid monitoring}\label{hybrid-monitoring}

In practice, a trial may benefit from running both a group sequential
boundary and an e-process on the same data, using the group sequential
rule as the primary decision criterion and the e-process as a
supplementary evidential summary. Table~\ref{tbl-hybrid} shows both
monitoring streams evaluated at each interim look for a single simulated
trial under \(H_1\).

\begin{longtable}[]{@{}
  >{\raggedleft\arraybackslash}p{(\linewidth - 18\tabcolsep) * \real{0.0633}}
  >{\raggedleft\arraybackslash}p{(\linewidth - 18\tabcolsep) * \real{0.0506}}
  >{\raggedleft\arraybackslash}p{(\linewidth - 18\tabcolsep) * \real{0.1392}}
  >{\raggedleft\arraybackslash}p{(\linewidth - 18\tabcolsep) * \real{0.1646}}
  >{\raggedleft\arraybackslash}p{(\linewidth - 18\tabcolsep) * \real{0.0759}}
  >{\raggedleft\arraybackslash}p{(\linewidth - 18\tabcolsep) * \real{0.1139}}
  >{\raggedright\arraybackslash}p{(\linewidth - 18\tabcolsep) * \real{0.1013}}
  >{\raggedleft\arraybackslash}p{(\linewidth - 18\tabcolsep) * \real{0.1139}}
  >{\raggedright\arraybackslash}p{(\linewidth - 18\tabcolsep) * \real{0.0886}}
  >{\raggedleft\arraybackslash}p{(\linewidth - 18\tabcolsep) * \real{0.0886}}@{}}

\caption{\label{tbl-hybrid}Hybrid monitoring table for a single
simulated trial under \(H_1\). At each interim look, both the e-process
and the group sequential boundary are evaluated. The always-valid
\(p\)-value is \(1/\max_{s \le n} E_s\).}

\tabularnewline

\toprule\noalign{}
\begin{minipage}[b]{\linewidth}\raggedleft
Look
\end{minipage} & \begin{minipage}[b]{\linewidth}\raggedleft
n
\end{minipage} & \begin{minipage}[b]{\linewidth}\raggedleft
Info frac.
\end{minipage} & \begin{minipage}[b]{\linewidth}\raggedleft
\(\hat\delta\)
\end{minipage} & \begin{minipage}[b]{\linewidth}\raggedleft
\(z\)
\end{minipage} & \begin{minipage}[b]{\linewidth}\raggedleft
GS bound
\end{minipage} & \begin{minipage}[b]{\linewidth}\raggedright
GS rej.
\end{minipage} & \begin{minipage}[b]{\linewidth}\raggedleft
\(\log E\)
\end{minipage} & \begin{minipage}[b]{\linewidth}\raggedright
E rej.
\end{minipage} & \begin{minipage}[b]{\linewidth}\raggedleft
AV \(p\)
\end{minipage} \\
\midrule\noalign{}
\endhead
\bottomrule\noalign{}
\endlastfoot
1 & 10 & 0.05 & 0.400 & 1.952 & 9.594 & FALSE & 0.985 & FALSE & 0.373 \\
2 & 20 & 0.10 & 0.350 & 2.394 & 6.784 & FALSE & 1.801 & FALSE & 0.165 \\
3 & 30 & 0.15 & 0.333 & 2.802 & 5.539 & FALSE & 2.514 & FALSE & 0.081 \\
4 & 40 & 0.20 & 0.325 & 3.128 & 4.797 & FALSE & 3.124 & FALSE & 0.036 \\
5 & 50 & 0.25 & 0.360 & 3.951 & 4.290 & FALSE & 4.484 & TRUE & 0.011 \\
6 & 60 & 0.30 & 0.333 & 4.038 & 3.917 & TRUE & 5.028 & TRUE & 0.007 \\
7 & 70 & 0.35 & 0.314 & 4.007 & 3.626 & TRUE & 5.366 & TRUE & 0.003 \\
8 & 80 & 0.40 & 0.262 & 3.576 & 3.392 & TRUE & 4.889 & TRUE & 0.003 \\
9 & 90 & 0.45 & 0.267 & 3.891 & 3.198 & TRUE & 5.704 & TRUE & 0.003 \\
10 & 100 & 0.50 & 0.270 & 4.131 & 3.034 & TRUE & 6.315 & TRUE & 0.002 \\
11 & 110 & 0.55 & 0.273 & 4.384 & 2.893 & TRUE & 7.130 & TRUE & 0.001 \\
12 & 120 & 0.60 & 0.258 & 4.299 & 2.769 & TRUE & 7.300 & TRUE & 0.000 \\
13 & 130 & 0.65 & 0.254 & 4.367 & 2.661 & TRUE & 7.638 & TRUE & 0.000 \\
14 & 140 & 0.70 & 0.286 & 5.118 & 2.564 & TRUE & 9.541 & TRUE & 0.000 \\
15 & 150 & 0.75 & 0.267 & 4.914 & 2.477 & TRUE & 9.130 & TRUE & 0.000 \\
16 & 160 & 0.80 & 0.269 & 5.125 & 2.398 & TRUE & 9.843 & TRUE & 0.000 \\
17 & 170 & 0.85 & 0.259 & 5.094 & 2.327 & TRUE & 9.910 & TRUE & 0.000 \\
18 & 180 & 0.90 & 0.250 & 5.040 & 2.261 & TRUE & 9.874 & TRUE & 0.000 \\
19 & 190 & 0.95 & 0.258 & 5.352 & 2.201 & TRUE & 10.961 & TRUE &
0.000 \\
20 & 200 & 1.00 & 0.240 & 5.089 & 2.145 & TRUE & 10.278 & TRUE &
0.000 \\

\end{longtable}

\begin{figure}[H]

\centering{

\includegraphics[width=0.8\linewidth,height=\textheight,keepaspectratio]{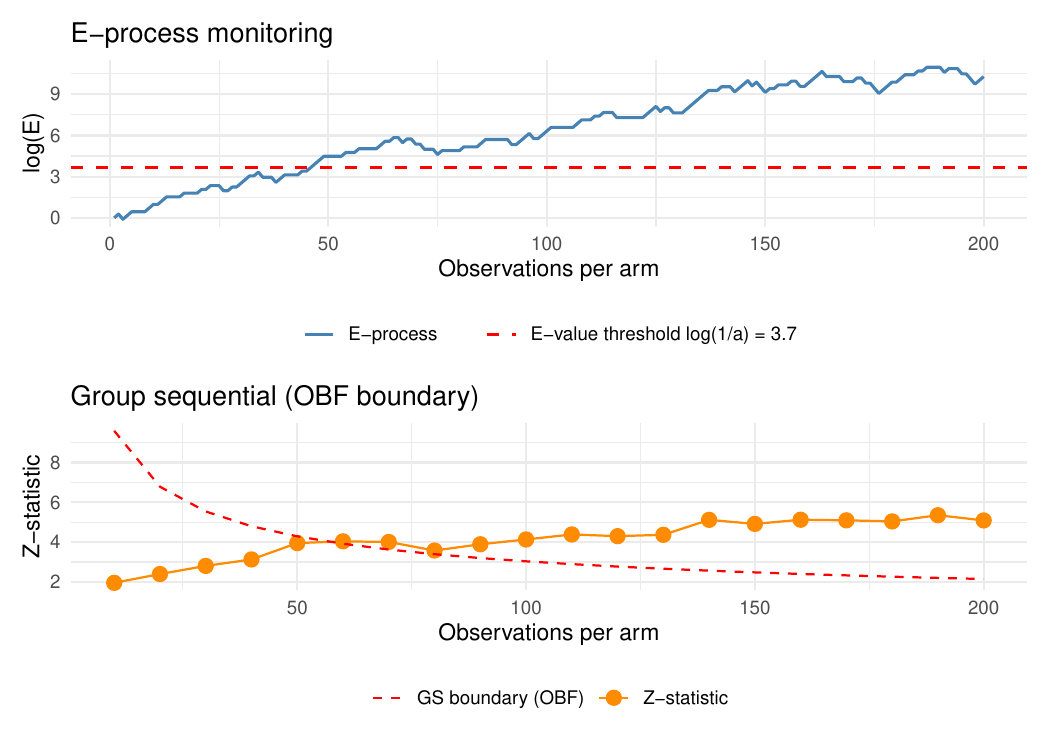}

}

\caption{\label{fig-hybrid}Hybrid monitoring on a single trial under
\(H_1\). Top panel: the log-e-process with its rejection threshold.
Bottom panel: the \(z\)-statistic at each look with the O'Brien--Fleming
boundary. Both streams cross their respective thresholds, but the timing
and trajectory differ.}

\end{figure}

Figure~\ref{fig-hybrid} shows the e-process and the group sequential
boundary on a shared time axis. The two monitoring streams provide
complementary information: the group sequential boundary yields a binary
signal at discrete looks, while the e-process provides a continuous
evidence trajectory.

Across the 50,000 simulated trials under \(H_1\), the group sequential
rule and the e-value rule agree in 72.0\% of trials (both reject) and
13.7\% of trials (neither rejects). In 14.1\% of trials the group
sequential rule rejects but the e-value does not, while in only 0.2\%
does the e-value reject without the group sequential rule also
rejecting. This concordance pattern confirms that e-value monitoring is
generally more conservative than calibrated group sequential monitoring,
but it provides a useful evidential summary that is valid regardless of
the monitoring schedule.

\subsection{Effect of monitoring schedule on the power
gap}\label{sec-irregular}

A key claim of e-value methodology is that its guarantees hold
regardless of the monitoring schedule. To test this operationally, we
compare e-value and recalibrated group sequential monitoring under four
look schedules: 5 fixed looks, 20 fixed looks, 5 randomly timed looks
(averaged over 200 random schedules), and continuous monitoring (every
patient pair). For each schedule, the OBF boundary constant \(c\) is
recalibrated by Monte Carlo so that the group sequential rule achieves
exactly \(\alpha = 0.025\) Type I error; the e-value threshold remains
\(1/\alpha = 40\) throughout.

\begin{longtable}[]{@{}
  >{\raggedright\arraybackslash}p{(\linewidth - 8\tabcolsep) * \real{0.3462}}
  >{\raggedright\arraybackslash}p{(\linewidth - 8\tabcolsep) * \real{0.2308}}
  >{\raggedleft\arraybackslash}p{(\linewidth - 8\tabcolsep) * \real{0.1667}}
  >{\raggedleft\arraybackslash}p{(\linewidth - 8\tabcolsep) * \real{0.0769}}
  >{\raggedleft\arraybackslash}p{(\linewidth - 8\tabcolsep) * \real{0.1795}}@{}}

\caption{\label{tbl-irregular}E-value vs recalibrated group sequential
power under four monitoring schedules. The e-value threshold is fixed at
\(1/\alpha\) regardless of schedule. The GS boundary is recalibrated for
each schedule. Under continuous monitoring, the GS boundary becomes
extremely conservative (power 10\%), while e-value power increases to
75\%.}

\tabularnewline

\toprule\noalign{}
\begin{minipage}[b]{\linewidth}\raggedright
Schedule
\end{minipage} & \begin{minipage}[b]{\linewidth}\raggedright
Method
\end{minipage} & \begin{minipage}[b]{\linewidth}\raggedleft
Type I error
\end{minipage} & \begin{minipage}[b]{\linewidth}\raggedleft
Power
\end{minipage} & \begin{minipage}[b]{\linewidth}\raggedleft
Avg. n (alt.)
\end{minipage} \\
\midrule\noalign{}
\endhead
\bottomrule\noalign{}
\endlastfoot
Fixed (5 looks) & E-value & 0.008 & 0.686 & 154.140 \\
Fixed (5 looks) & GS (recalibrated) & 0.025 & 0.870 & 149.738 \\
Fixed (20 looks) & E-value & 0.012 & 0.723 & 139.230 \\
Fixed (20 looks) & GS (recalibrated) & 0.025 & 0.861 & 139.819 \\
Irregular (5 random looks) & E-value & 0.007 & 0.679 & 160.365 \\
Irregular (5 random looks) & GS (recalibrated) & 0.025 & 0.868 &
157.669 \\
Continuous (200 looks) & E-value & 0.016 & 0.750 & 131.056 \\
Continuous (200 looks) & GS (recalibrated) & 0.043 & 0.100 & 180.192 \\

\end{longtable}

\begin{figure}[H]

\centering{

\includegraphics[width=0.7\linewidth,height=\textheight,keepaspectratio]{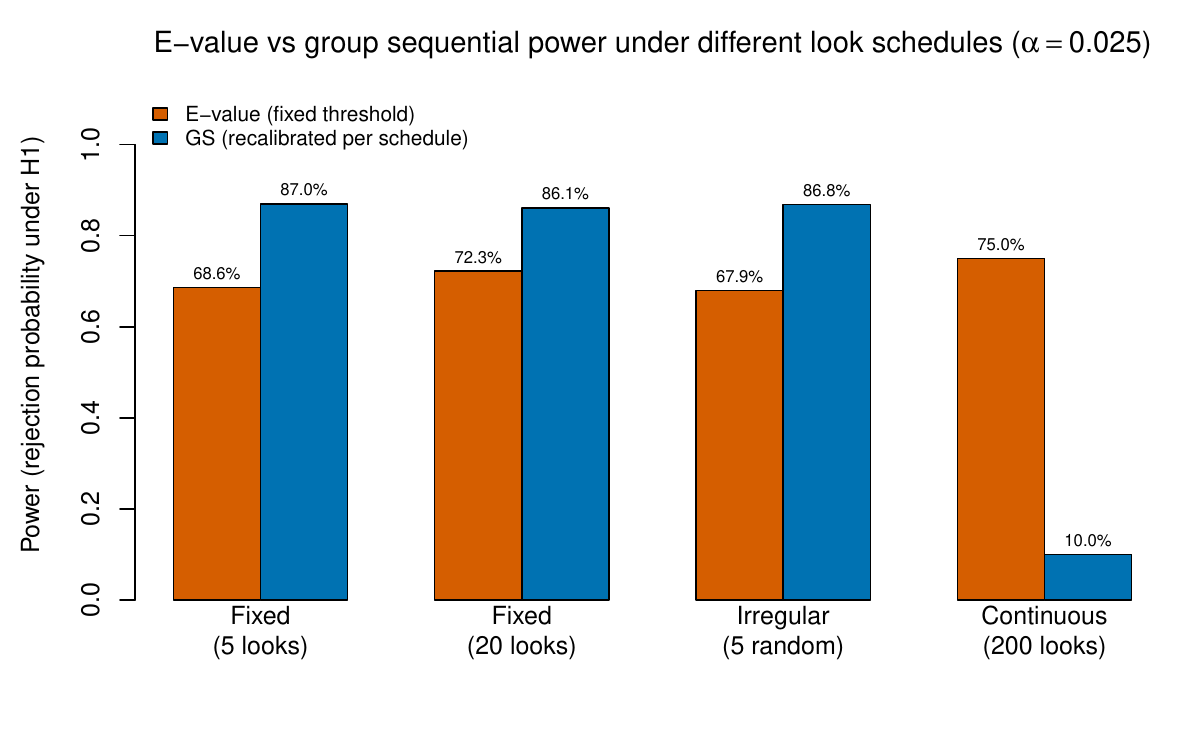}

}

\caption{\label{fig-irregular}E-value vs group sequential power under
four monitoring schedules. The e-value threshold is fixed; the GS
boundary is recalibrated for each schedule. Under continuous monitoring,
the GS power collapses while e-value power increases.}

\end{figure}

The results in Table~\ref{tbl-irregular} show a clear pattern. Under
fixed schedules with few looks, the group sequential rule substantially
outperforms the e-value (87\% vs 69\% power with 5 looks). As the number
of looks increases to 20, the GS advantage narrows slightly (86\% vs
72\%). But under continuous monitoring---the regime where e-values are
designed to excel---the GS boundary, even after recalibration, achieves
only 10.0\% power, while the e-value achieves 75.0\%. The explanation is
that the OBF boundary \(z \ge c/\sqrt{t}\) must be inflated to maintain
Type I error under many looks, eventually becoming so conservative at
early looks that the boundary is effectively unreachable. The e-value
threshold, by contrast, remains unchanged. Under irregular scheduling
(random look times), the e-value maintains stable power while the GS
rule requires re-calibration for each realized schedule---an operational
complication that e-values avoid entirely. Figure~\ref{fig-irregular}
visualizes these results.

\subsection{Parameter sensitivity across design
settings}\label{parameter-sensitivity-across-design-settings}

The power gap between e-values and group sequential methods is not
constant: it depends on the base rate \(p_C\), the effect size
\(\delta\), and the monitoring intensity. Table~\ref{tbl-grid} reports
e-value and GS power across a grid of 18 design configurations
(\(p_C \in \{0.10, 0.30, 0.50\}\), \(\delta \in \{0.10, 0.15, 0.20\}\),
\(K \in \{5, 20\}\) looks), each based on 20,000 Monte Carlo
replications.

\begin{longtable}[]{@{}rrrrrrr@{}}

\caption{\label{tbl-grid}E-value and group sequential power across
design configurations (\(N_{\max} = 200\) per arm, \(\alpha = 0.025\)).
The power gap is largest for small effects at moderate base rates and
smallest for large effects at low base rates.}

\tabularnewline

\toprule\noalign{}
\(p_C\) & \(\delta\) & Looks & \(\lambda^*\) & E-value & GS & Gap \\
\midrule\noalign{}
\endhead
\bottomrule\noalign{}
\endlastfoot
0.1 & 0.10 & 5 & 0.385 & 0.583 & 0.795 & 0.211 \\
0.3 & 0.10 & 5 & 0.217 & 0.262 & 0.547 & 0.285 \\
0.5 & 0.10 & 5 & 0.200 & 0.223 & 0.499 & 0.276 \\
0.1 & 0.15 & 5 & 0.500 & 0.891 & 0.977 & 0.086 \\
0.3 & 0.15 & 5 & 0.312 & 0.685 & 0.864 & 0.179 \\
0.5 & 0.15 & 5 & 0.300 & 0.659 & 0.840 & 0.181 \\
0.1 & 0.20 & 5 & 0.588 & 0.978 & 0.999 & 0.021 \\
0.3 & 0.20 & 5 & 0.400 & 0.900 & 0.980 & 0.080 \\
0.5 & 0.20 & 5 & 0.400 & 0.900 & 0.978 & 0.077 \\
0.1 & 0.10 & 20 & 0.385 & 0.619 & 0.785 & 0.166 \\
0.3 & 0.10 & 20 & 0.217 & 0.292 & 0.533 & 0.241 \\
0.5 & 0.10 & 20 & 0.200 & 0.250 & 0.494 & 0.244 \\
0.1 & 0.15 & 20 & 0.500 & 0.908 & 0.974 & 0.066 \\
0.3 & 0.15 & 20 & 0.312 & 0.718 & 0.856 & 0.139 \\
0.5 & 0.15 & 20 & 0.300 & 0.696 & 0.844 & 0.149 \\
0.1 & 0.20 & 20 & 0.588 & 0.983 & 0.999 & 0.016 \\
0.3 & 0.20 & 20 & 0.400 & 0.918 & 0.979 & 0.060 \\
0.5 & 0.20 & 20 & 0.400 & 0.917 & 0.979 & 0.062 \\

\end{longtable}

Two patterns emerge. First, the power gap narrows substantially as the
effect size increases: for \(\delta = 0.20\) with \(p_C = 0.10\), the
gap is only 2.1 percentage points (e-value 97.8\% vs GS 99.9\%), while
for \(\delta = 0.10\) with \(p_C = 0.30\) it widens to 28.5 percentage
points. Second, increasing the number of looks from 5 to 20 narrows the
gap modestly (typically by 3--5 percentage points), because more
frequent monitoring allows the e-value to detect threshold crossings
earlier while the GS boundary must spread its alpha budget more thinly.
The base rate \(p_C\) also matters: lower base rates (more discordant
pairs per patient pair) yield faster growth rates and higher power for
both methods, with the e-value benefiting relatively more.

\subsection{Futility monitoring
demonstration}\label{futility-monitoring-demonstration}

To illustrate the futility monitoring methodology described in Section
Section~\ref{sec-eprocesses}, we simulate a scenario where the treatment
has a small but clinically irrelevant effect: \(p_T = 0.33\),
\(p_C = 0.3\) (true \(\delta = 0.03\)), with a minimum clinically
important difference of \(\delta_{\min} = 0.1\). The maximum sample size
is 300 per arm.

We apply both futility routes: (i) the confidence-sequence approach,
which declares futility when the upper bound of the 95\% CS falls below
\(\delta_{\min}\), and (ii) the reciprocal e-process, which tests
\(H_0'{:}\, \delta \ge \delta_{\min}\) at \(\alpha_f = 0.10\). Over
10,000 simulations, the CS-based method detects futility in 14.2\% of
trials (median detection at 147.5 pairs among those detected), while the
reciprocal e-process detects futility in 53.5\% of trials (median at 108
pairs). The reciprocal e-process is more sensitive because it directly
tests the futility hypothesis, while the CS route requires the entire
confidence band to exclude \(\delta_{\min}\)---a more conservative
criterion. Both mechanisms are implemented in \texttt{evalinger}
(\texttt{futility\_cs()} and \texttt{futility\_eprocess()}).

\begin{figure}[H]

\centering{

\includegraphics[width=1\linewidth,height=\textheight,keepaspectratio]{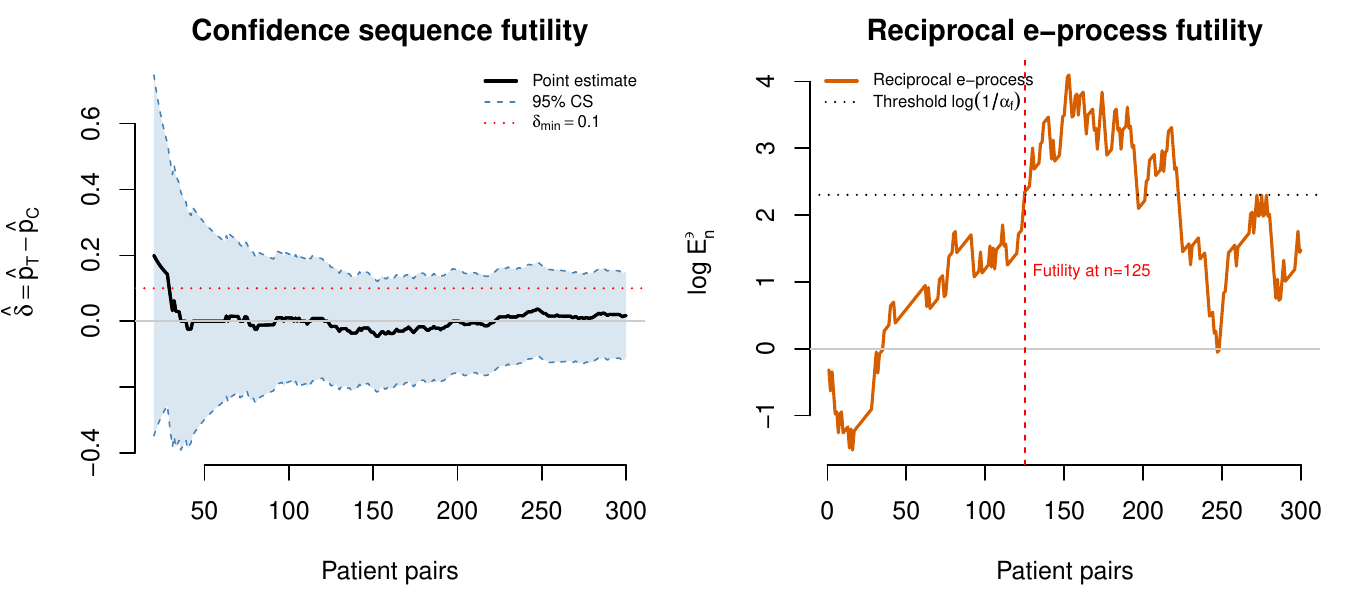}

}

\caption{\label{fig-futility}Futility monitoring for a trial with a
subclinical effect (\(p_T = 0.33\), \(p_C = 0.30\),
\(\delta_{\min} = 0.10\)). Left: the 95\% confidence sequence with the
MCID threshold; futility is declared when the upper bound falls below
\(\delta_{\min}\). Right: the reciprocal e-process testing
\(H_0'{:}\, \delta \ge \delta_{\min}\); futility is declared when the
process crosses \(\log(1/\alpha_f)\).}

\end{figure}

Figure~\ref{fig-futility} illustrates both approaches on a single trial.
The left panel shows the confidence sequence narrowing around the true
effect (\(\delta = 0.03\)) until its upper bound drops below
\(\delta_{\min} = 0.10\), triggering futility. The right panel shows the
reciprocal e-process growing as evidence accumulates against a
clinically meaningful effect.

\subsection{RECOVERY-like large-trial
simulation}\label{sec-recovery-sim}

To demonstrate the methodology at a scale typical of modern confirmatory
trials, we simulate a trial modeled on the RECOVERY dexamethasone arm
(RECOVERY Collaborative Group 2021). The published results reported
28-day mortality of 22.9\% (dexamethasone) vs 25.7\% (usual care), an
absolute reduction of 2.8 percentage points. We simulate a 1:1
randomized trial with \(N = 2000\) per arm using these rates.

The GROW-optimal betting fraction is \(\lambda^* = 0.0760\), with growth
rate \(g(\lambda^*) = 0.001065\) nats per pair---an order of magnitude
slower than the main study (\(g \approx 0.024\)), reflecting the much
smaller effect size. The expected stopping time is approximately 3462
pairs, exceeding the 2000-per-arm trial. In a single simulated trial,
the e-process reaches \(E = 8.8\) (below the threshold of
\(1/\alpha = 40\)), with an always-valid 95\% confidence sequence of
\([-0.019,\, 0.075]\) for the mortality difference. Over 10,000
simulations, the e-value achieves 31.4\% power at \(N = 2000\) per arm,
with median rejection at 1417 pairs among trials that reject.

\begin{figure}[H]

\centering{

\includegraphics[width=1\linewidth,height=\textheight,keepaspectratio]{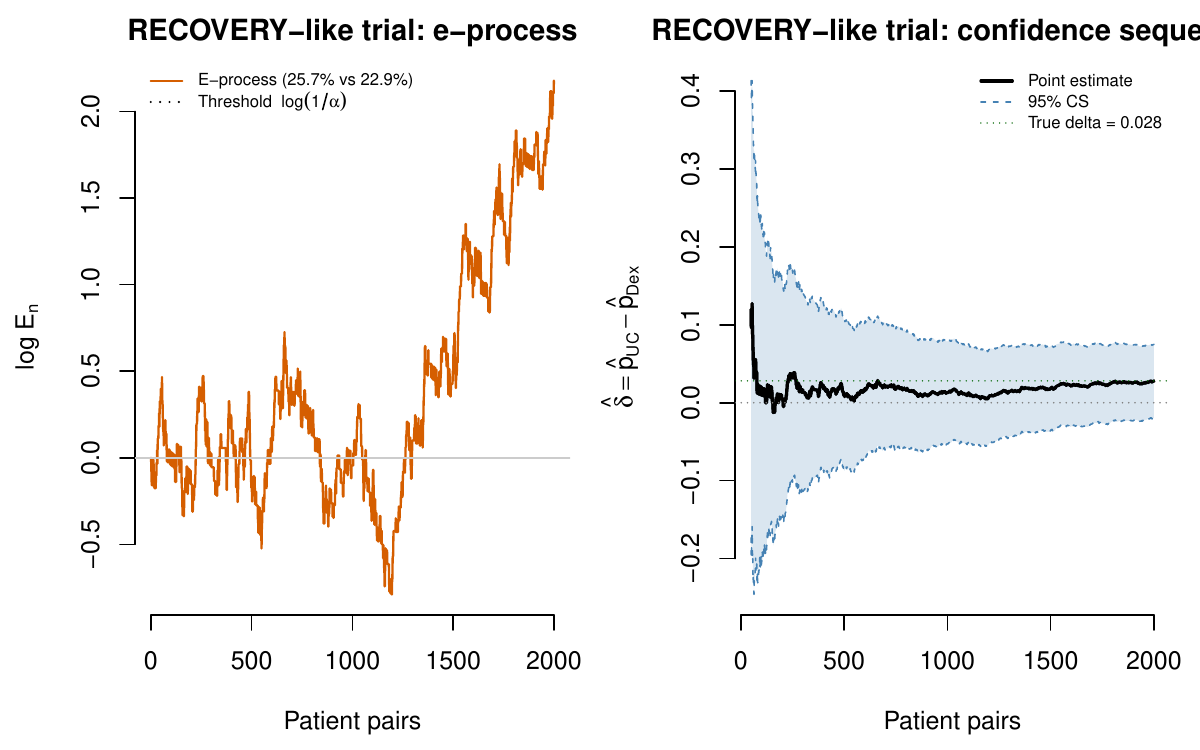}

}

\caption{\label{fig-recovery}RECOVERY-like trial simulation
(\(p_{UC} = 25.7\%\), \(p_{Dex} = 22.9\%\), \(N = 2000\) per arm). Left:
the e-process trajectory grows steadily but does not reach the rejection
threshold within 2000 pairs, reflecting the small effect size. Right:
the 95\% confidence sequence narrows around the true effect and provides
anytime-valid estimation.}

\end{figure}

This analysis illustrates two important points. First, for small effect
sizes (\(\delta \approx 0.03\)), the e-value requires very large samples
to achieve adequate power---the expected stopping time of 3462 pairs
substantially exceeds what a typical trial can enroll. This is the
regime where group sequential methods, which can optimize boundaries for
a fixed schedule, have the largest advantage. Second, the confidence
sequence provides useful information even when the e-process does not
cross the rejection threshold: the interval \([-0.019,\, 0.075]\) is
anytime-valid and correctly contains the true effect, providing an
interpretable summary of the evidence that a DSMB could use alongside
group sequential decisions. Figure~\ref{fig-recovery} shows the
e-process trajectory and confidence sequence.

\section{Real-data illustration: the Novick (1965) ulcer
trial}\label{sec-novick}

To illustrate the methodology on real clinical trial data, we revisit
one of the earliest Bayesian analyses of a randomized controlled trial:
Novick and Grizzle's 1965 study comparing four surgical treatments for
duodenal ulcer (Novick and Grizzle 1965). The trial, conducted across 19
Veterans Administration hospitals, randomized 400 patients (100 per arm)
to treatments A, B, C, and D. Outcomes at six months were classified as
excellent/good, fair/poor, or death. The observed death counts were: A =
7, B = 1, C = 1, D = 3.

This dataset is well suited for our purposes for three reasons. First,
the original paper analyzed the data using \emph{both} a classical
restricted sequential design and a Bayesian posterior credibility
analysis, providing a natural three-way comparison when we add the
e-value perspective. Second, the multi-arm structure allows us to
demonstrate platform monitoring and multiplicity control via
\texttt{platform\_monitor()} and \texttt{ebh()}. Third, Novick himself
argued that the Bayesian approach provided ``a degree of flexibility not
afforded by classical methods'' for sequential re-evaluation---precisely
the kind of flexibility that e-values formalize with time-uniform
guarantees.

\subsection{E-process monitoring}\label{e-process-monitoring}

We treat the death outcome as a binary endpoint and ask whether each
treatment has a higher mortality rate than Treatment B (the best
performer). For each pairwise comparison, we construct a
betting-martingale e-process using \texttt{eprocess\_binary()} with a
common GROW-optimal betting fraction \(\lambda^* = 0.727\), calibrated
for a design alternative of 6\% vs 1\% mortality (informed by the
observed rates in the trial; Novick's prior expectation of 2--3\%
baseline mortality is broadly compatible with these values). The
significance level is \(\alpha = 0.025\) (one-sided).

\begin{figure}[H]

\centering{

\includegraphics[width=0.85\linewidth,height=\textheight,keepaspectratio]{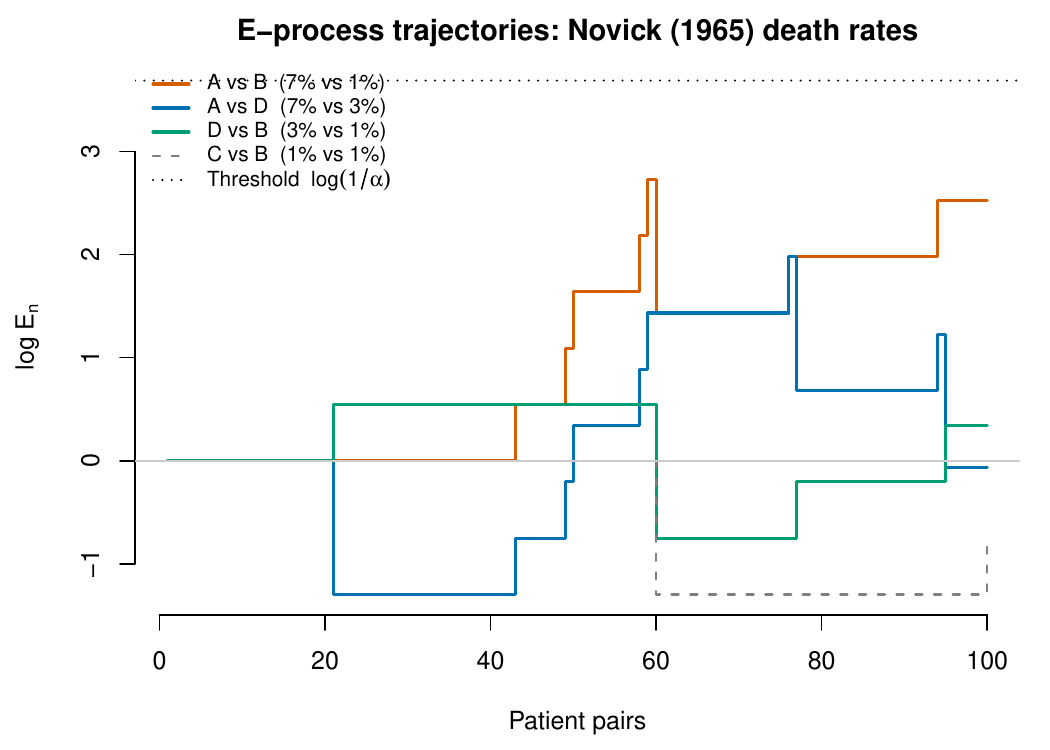}

}

\caption{\label{fig-novick-eprocess}E-process trajectories for four
pairwise comparisons of death rates in the Novick (1965) trial. The
horizontal dashed line is the rejection threshold
\(\log(1/\alpha) = 3.69\). The strongest comparison (A vs B, 7\% vs 1\%)
accumulates substantial evidence (\(E \approx 15.3\)) but does not cross
the threshold, illustrating the conservatism of anytime-valid inference
with 100 patients per arm and rare events. The C vs B comparison (both
1\%) correctly shows no evidence accumulation.}

\end{figure}

Since the original patient-level sequential ordering is not available,
we simulate a random arrival order within each arm (with a fixed seed
for reproducibility) and construct the e-process as each patient pair is
observed. Figure~\ref{fig-novick-eprocess} shows the resulting evidence
trajectories.

Table~\ref{tbl-novick} compares the e-value evidence with Novick's
original Bayesian posterior credibilities. Novick's Bayesian analysis
reported credibility \(C(p_A > p_B) \approx 0.98\) for the A vs B
comparison, suggesting very strong evidence of a mortality difference.
The e-process, by contrast, yields \(E \approx 15.3\), corresponding to
an always-valid \(p\)-value of 0.065---moderate evidence that does not
reach the \(\alpha = 0.025\) threshold. This difference reflects the
price of anytime validity: the e-value guarantee holds regardless of how
many times the data are inspected, whereas the Bayesian credibility does
not automatically control the probability of ever declaring a false
positive under repeated monitoring. With the design alternative used
here, the expected stopping time is approximately 182 patient
pairs---more than the 100 available---confirming that the trial is
underpowered for definitive e-value rejection at this \(\alpha\) level.

\begin{longtable}[]{@{}
  >{\raggedright\arraybackslash}p{(\linewidth - 10\tabcolsep) * \real{0.1571}}
  >{\raggedleft\arraybackslash}p{(\linewidth - 10\tabcolsep) * \real{0.2429}}
  >{\raggedleft\arraybackslash}p{(\linewidth - 10\tabcolsep) * \real{0.2429}}
  >{\raggedleft\arraybackslash}p{(\linewidth - 10\tabcolsep) * \real{0.0714}}
  >{\raggedleft\arraybackslash}p{(\linewidth - 10\tabcolsep) * \real{0.1000}}
  >{\raggedleft\arraybackslash}p{(\linewidth - 10\tabcolsep) * \real{0.1857}}@{}}

\caption{\label{tbl-novick}Pairwise comparison of death rates in the
Novick (1965) trial. \(E\) is the maximum e-value attained over the
sequential monitoring period; AV \(p\) is the always-valid \(p\)-value
(\(1/\max E\)); Novick's credibility is the posterior probability
\(C(\lambda_i/\lambda_j > 1)\) under a Poisson model with uniform prior
(from Novick's Table 4).}

\tabularnewline

\toprule\noalign{}
\begin{minipage}[b]{\linewidth}\raggedright
Comparison
\end{minipage} & \begin{minipage}[b]{\linewidth}\raggedleft
Deaths (arm \(i\))
\end{minipage} & \begin{minipage}[b]{\linewidth}\raggedleft
Deaths (arm \(j\))
\end{minipage} & \begin{minipage}[b]{\linewidth}\raggedleft
\(E\)
\end{minipage} & \begin{minipage}[b]{\linewidth}\raggedleft
AV \(p\)
\end{minipage} & \begin{minipage}[b]{\linewidth}\raggedleft
Novick cred.
\end{minipage} \\
\midrule\noalign{}
\endhead
\bottomrule\noalign{}
\endlastfoot
A vs B & 7 & 1 & 15.3 & 0.065 & 0.980 \\
A vs C & 7 & 1 & 45.8 & 0.022 & 0.980 \\
A vs D & 7 & 3 & 7.2 & 0.138 & 0.887 \\
D vs B & 3 & 1 & 1.7 & 0.579 & 0.813 \\
D vs C & 3 & 1 & 5.2 & 0.194 & 0.813 \\

\end{longtable}

\subsection{Confidence sequence and platform
monitoring}\label{confidence-sequence-and-platform-monitoring}

\begin{figure}[H]

\centering{

\includegraphics[width=0.8\linewidth,height=\textheight,keepaspectratio]{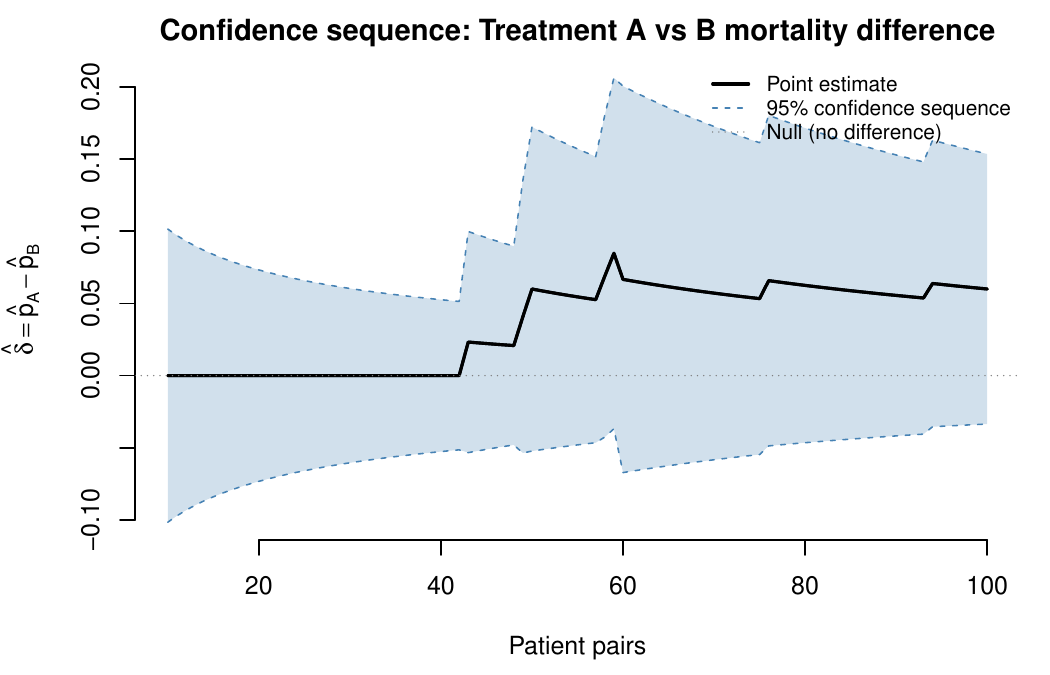}

}

\caption{\label{fig-novick-confseq}Always-valid 95\% confidence sequence
for the mortality difference \(\delta = p_A - p_B\) in the Novick (1965)
trial. The confidence band shrinks as data accumulate but still includes
zero at \(n = 100\), consistent with the non-rejection of the
e-process.}

\end{figure}

The always-valid confidence sequence for the treatment effect
\(\delta = p_A - p_B\) at \(n = 100\) is \([-0.034,\, 0.154]\) with
point estimate \(\hat{\delta} = 0.06\)
(Figure~\ref{fig-novick-confseq}). Because this interval includes zero,
the e-value analysis is consistent: the data do not provide
anytime-valid evidence to exclude the possibility of equal mortality
rates, even though the point estimate suggests a 6 percentage-point
difference.

When the trial is analyzed as a platform (Treatment B as shared control,
arms A, C, D as experimental), the e-BH procedure with FDR level 0.05
does not reject any hypothesis at any of the four interim looks
(\(n = 25, 50, 75, 100\) per arm). This illustrates the additional
conservatism of multiplicity control and confirms that 100 patients per
arm with 1--7\% event rates is insufficient for multiplicity-adjusted
rejection---an appropriate conclusion given the limited data.

The three-way comparison across analytic paradigms is informative.
Novick's classical sequential analysis (Armitage's restricted sequential
plan) also failed to reach a decision boundary after 400 patients: the
preference-based test showed no significant pairwise difference for any
of the six comparisons, and Novick estimated that approximately 200
additional plottable pairs would be needed before the A-versus-B
comparison might reach a boundary. The e-value analysis agrees with this
frequentist conclusion: \(E \approx 15.3\) is substantial evidence but
insufficient for rejection at \(\alpha = 0.025\). Novick's Bayesian
credibility of 0.98, by contrast, is the outlier---it suggests
near-certainty, but this high credibility does not carry a frequentist
error-control guarantee under optional stopping. This is consistent with
the theoretical framework developed here: e-values are a frequentist
tool and agree with classical methods on what constitutes ``enough
evidence,'' while providing operational advantages under flexible
monitoring. Unlike the classical sequential test, which could only
declare ``no significant difference'' at its pre-planned boundary
crossings, the e-value analysis produces a continuous evidence
trajectory (\(E = 15.3\), corresponding to an always-valid \(p\)-value
of 0.065), a shrinking confidence sequence that is valid at every sample
size, and a guarantee that these summaries remain valid regardless of
how many times the investigators examined the accumulating data---the
very flexibility that Novick argued was exclusive to the Bayesian
approach. The code for this analysis is available in
\texttt{run\_novick\_analysis.R} and uses the \texttt{evalinger}
functions \texttt{eprocess\_binary()}, \texttt{confseq\_binary()},
\texttt{platform\_monitor()}, and \texttt{ebh()}.

\section{Discussion}\label{discussion}

E-values provide a time-uniform evidential interface suited to the
sequential nature of clinical trials. Compared with alpha-spending and
group sequential monitoring, e-process monitoring simplifies governance
when interim looks are irregular or opportunistic and provides a
coherent mechanism for optional continuation. Compared with Bayesian
posterior or predictive-probability monitoring, e-values offer a direct
frequentist guarantee under optional stopping without extensive
simulation-based calibration, while still permitting Bayesian-inspired
constructions (mixtures, priors) to improve efficiency.

The numerical study reveals a power gap of approximately 14 percentage
points between the calibrated group sequential boundary (86.1\%) and the
betting e-process (72.3\%) under 20 equally spaced looks. However, the
extended simulations (Sections
Section~\ref{sec-irregular}--Section~\ref{sec-recovery-sim}) show that
this gap is highly context-dependent. Under continuous monitoring, the
GS power collapses to 10\% while e-value power increases to 75\%,
because the OBF boundary must become extremely conservative to maintain
Type I error under many looks. The parameter grid (Table~\ref{tbl-grid})
shows the gap ranges from 2 percentage points (large effects) to 29
percentage points (small effects at moderate base rates), and the
RECOVERY-like simulation (Section~\ref{sec-recovery-sim}) confirms that
for small effects (\(\delta \approx 0.03\)), e-values require sample
sizes substantially exceeding typical trial enrollment. Several
strategies can narrow the gap under fixed schedules: adaptive betting
fractions \(\lambda_i\) that incorporate accumulating information,
regularized e-processes (Martin 2025), and hybrid approaches using a GS
boundary at pre-planned looks and an e-process for unplanned looks.

The calibrated Bayesian monitoring rule is calibrated to control Type I
error (by construction) but is not automatically more powerful than an
e-value design once calibration is imposed. In our example, the
calibrated Bayesian rule achieves Type I error 2.0\% with power 68.8\%
and average stopping of 134 per arm under the alternative. The
disadvantage is that calibration is simulation-specific: the posterior
threshold of approximately 0.998 was computed for a particular null
hypothesis (\(p_T = p_C = 0.30\)), look schedule (20 looks), and
analysis prior (Jeffreys \(\text{Beta}(0.5,0.5)\) on each arm); changing
any of these requires re-simulation. This brittleness is a practical
cost of posterior-threshold monitoring, and it contrasts with the
e-value threshold \(1/\alpha\), which remains valid under optional
stopping without re-calibration once a valid e-process has been
specified. The 2026 FDA draft Bayesian guidance (U.S. Food and Drug
Administration 2026) illuminates this tradeoff from a regulatory
perspective by distinguishing \emph{analysis priors} from \emph{design
priors}: the calibration threshold is tied to the analysis prior, while
the power assessments in our Table~\ref{tbl-main} correspond to
operating characteristics under a design alternative. The guidance
requires that operating characteristics be evaluated under "various
plausible design priors," which means that calibration must be revisited
across a range of null and alternative scenarios---a simulation burden
that grows with trial complexity. By contrast, e-values provide their
Type I error guarantee analytically for any stopping rule; the design
task shifts primarily to power optimization via choice of the betting
fraction and to sensitivity analysis (as in
Table~\ref{tbl-sensitivity}).

E-values are not a universal replacement for classical tools. Their
practical performance depends on the chosen betting strategy, and poorly
chosen e-processes can be severely underpowered. For complex
endpoints---such as time-to-event outcomes with non-proportional
hazards, recurrent events, or longitudinal outcomes with informative
missingness---constructing powerful yet robust e-processes remains a
research frontier. Berry's experience illustrates this: in the Pravigard
Pac meta-analysis, the investigators deliberately used non-proportional
hazards models because ``one of the two agents might carry the combo's
efficacy early on, say in the first year, with the other agent taking
over for the later term'' (S. M. Berry et al. 2004; D. A. Berry 2025),
and in GBM AGILE and Precision Promise, non-proportional hazard models
are used routinely for overall survival analysis. The safe logrank
e-process described in Section Section~\ref{sec-safe-logrank} assumes
proportional hazards, and extending it to non-proportional settings is
an important open problem. Similarly, Berry's use of longitudinal models
to predict late endpoints from early measurements---a technique that
saved ``at least a year of development time'' in AWARD-5 by enabling a
seamless phase shift before any patient had reached the defining
12-month visit (D. A. Berry 2025)---points to a practical capability
that e-process methodology does not yet possess but would substantially
increase its operational value. Composite nulls with high-dimensional
nuisance parameters (e.g., covariate-adjusted treatment effects) require
either conditioning, which may sacrifice efficiency, or worst-case
constructions, which may be conservative. The interaction between
e-values and response-adaptive randomization in high-dimensional
covariate settings, where both the filtration and the estimand may
evolve over time, is not yet well understood. More broadly,
quantile-based prediction methods that learn predictive distributions
directly (Nareklishvili, Polson, and Sokolov 2025) offer a complementary
route to calibration-free inference that may connect to e-value
methodology through conformal e-prediction (Polson, Sokolov, and
Zantedeschi 2026; Datta et al. 2025).

From a regulatory perspective, the gap between the e-value literature
and established regulatory templates is narrowing. The 2026 FDA draft
Bayesian guidance (U.S. Food and Drug Administration 2026) does not
mention e-values, but its three-pathway framework for success
criteria---calibrated to Type I error rate, direct posterior
interpretation, and decision-theoretic---creates natural entry points
for e-value methodology (see Section Section~\ref{sec-regulatory} for
detailed discussion). The mapping to familiar reporting quantities
(always-valid \(p\)-values, confidence sequences) and the inclusion of
simulation-based operating characteristics in the statistical analysis
plan remain essential steps toward regulatory acceptance, and the
guidance's requirement for comprehensive operating characteristics
documentation applies equally to e-value-based designs.

An additional limitation is conceptual: e-values are an evidential scale
and not a complete decision theory. In confirmatory trials, decisions
are constrained by clinical context, safety considerations,
multiplicity, and regulatory requirements. E-values provide time-uniform
error control, but they do not remove the need to specify which
decisions are permissible at interim looks, which populations are
affected by adaptations, and which effect measures will be reported.

The RECOVERY-like simulation in Section Section~\ref{sec-recovery-sim}
demonstrates that e-value methodology scales to large modern trials, but
also highlights a practical limitation: when the treatment effect is
small (\(\delta \approx 0.03\)), the per-observation growth rate is
correspondingly small, and the expected stopping time exceeds what most
trials can enroll. A full-scale retrospective e-value analysis of the
actual RECOVERY trial, using patient-level data accessible through the
trial team's data-sharing mechanism, would provide a benchmark for
comparing e-value, group sequential, and Bayesian monitoring in a trial
where the interim decision materially affected patient care (RECOVERY
Collaborative Group 2021).

\section{Conclusion}\label{conclusion}

E-values and e-processes offer a principled framework for anytime-valid
inference under optional stopping and continuation, making them a
natural candidate for interim monitoring in adaptive clinical trials.
Their main advantages are explicit time-uniform Type I error control
without prespecifying a stopping time, coherent evidence accumulation
across looks and (under conditions) across studies, natural handling of
composite null hypotheses through the betting construction, and
compatibility with adaptive decision-making when design changes are
predictable functions of past information. Their main disadvantages are
a power gap relative to optimally calibrated group sequential boundaries
when the monitoring schedule is fixed, dependence of efficiency on the
choice and tuning of the e-process, additional methodological work
needed for complex endpoints and high-dimensional nuisance parameters,
and the current gap between the e-value literature and established
regulatory templates.

The numerical study on a two-arm binary trial provides concrete
quantification: among error-controlling methods, the calibrated GS
boundary achieves the highest power (86.1\%) under a fixed 20-look
schedule, the calibrated Bayesian rule is more conservative (68.8\%),
and the e-value provides robust anytime-valid control (72.3\%). The
extended simulations show this ranking reverses under continuous
monitoring (e-value 75\%, GS 10\%), and the power gap varies from 2 to
29 percentage points depending on the effect size and base rate. The
futility demonstration confirms that e-processes provide effective
futility monitoring, and the RECOVERY-like simulation shows the
methodology scales to large trials while highlighting the sample-size
cost of small effects. These findings motivate hybrid approaches in
which e-values complement, rather than replace, group sequential and
Bayesian methods.

For practitioners, a concise recommendation is as follows. E-values are
most appropriate when a trial will realistically face flexible
monitoring or data-dependent continuation and adaptation, and when an
auditable validity guarantee under those operations is a first-order
requirement---precisely the settings that Berry's platform trials (I-SPY
2, GBM AGILE, REMAP-CAP) have shown to be increasingly common (D. A.
Berry 2025). Conventional group sequential designs are preferable when
the monitoring plan is stable and regulatory-facing power efficiency
under standard assumptions is paramount. Bayesian predictive
probabilities, as Berry has demonstrated across numerous landmark
trials, are natural when the operational decision is explicitly about
the probability of future trial success and when algorithmic adaptation
should drive enrollment and allocation in real time; in such designs,
thresholds should be calibrated for frequentist operating
characteristics when such guarantees are required. In practice, the
three frameworks are most effective in combination: Bayesian
predictive-probability algorithms for operational decision-making, group
sequential boundaries for power-optimized confirmatory analysis at
pre-planned looks, and e-processes as an anytime-valid evidential ledger
that remains valid regardless of how many times the data are inspected,
how the schedule changes, or whether the trial continues beyond its
planned horizon.

The regulatory landscape favors such integration. The 2026 FDA draft
Bayesian guidance (U.S. Food and Drug Administration 2026) opens three
pathways for success criteria---Type I error calibration, direct
posterior interpretation, and decision-theoretic approaches---each of
which can accommodate e-value monitoring as a time-uniform evidential
layer.

The \texttt{evalinger} R package
(\url{https://github.com/VadimSokolov/evalinger}) and its interactive
web application (\url{https://sailtargets.shinyapps.io/evalinger/})
implement the full methodology and complement the broader treatment of
adaptive trial design in Sokolova and Sokolov (2026). The package
provides GROW-optimal design calibration, real-time monitoring,
confidence sequences, futility analysis, platform trial multiplicity
control, and head-to-head comparison with group sequential and Bayesian
rules. The web application's three dashboards (Design Calculator,
Monitoring Dashboard, and Method Comparison) generate the operating
characteristics documentation that the 2026 FDA draft guidance requires
(U.S. Food and Drug Administration 2026): Type I error rates, power
curves, expected sample sizes under design priors, and sensitivity to
calibration parameters.

\phantomsection\label{refs}

\end{document}